\documentclass[superscriptaddress,preprint, twocolumn,10pt, prx]{revtex4-2}%
\usepackage{amsmath}
\usepackage{amsfonts}
\usepackage{amssymb}
\usepackage{graphicx}
\usepackage[caption=false]{subfig}
\graphicspath{{pics/}}
\usepackage{aps_math}
\usepackage{basic_phys}
\usepackage{basic_comment}

\newcommand{\wjy}[1]{#1}
\newcommand{\gb}[1]{#1}
\usepackage{ifthen}
\def\bibfile{DQC_QV_ref}% for local bibfile
\ifthenelse{\equal{\bibfile}{}}{%
  \def\myprintbibliography{}%
}{%
  \def\myprintbibliography{%
    \bibliographystyle{qiounsrt}%
    \bibliography{\bibfile}%
  }%
}%
\usepackage[%
]{hyperref}
\def\myprintglossary{%
}
\begin{document}
%%%%%%%%%%%%%%%%%%%%%%%%%%%%%%%%%%%%%%%%%%%%%%%%%%%%%%%%%%%%%%%%%%%%%%
%%                         Title page                               %%
%%                                                                  %%
\title{Scalability enhancement of quantum computing under limited connectivity \\through distributed quantum computing}
\author{Shao-Hua Hu}
\affiliation{Department of Physics, Tamkang University, New Taipei 25137, Taiwan, ROC}

\author{George Biswas}
\affiliation{Department of Physics, Tamkang University, New Taipei 25137, Taiwan, ROC}
\affiliation{Center for Advanced Quantum Computing, Tamkang University, New Taipei 25137, Taiwan, ROC}

\author{Jun-Yi Wu}
\email{junyiwuphysics@gmail.com}
\affiliation{Department of Physics, Tamkang University, New Taipei 25137, Taiwan, ROC}
\affiliation{Center for Advanced Quantum Computing, Tamkang University, New Taipei 25137, Taiwan, ROC}
\affiliation{Physics Division, National Center for Theoretical Sciences, Taipei 10617, Taiwan, ROC}

\begin{abstract}
  We employ quantum-volume random-circuit sampling to benchmark the two-QPU entanglement-assisted distributed quantum computing (DQC) and compare it with single-QPU quantum computing. We first specify a single-qubit depolarizing noise model in the random circuit. Based on this error model, we show the one-to-one correspondence of three figures of merits, namely average gate fidelity, heavy output probability, and linear cross-entropy. We derive an analytical approximation of the average gate fidelity under the specified noise model, which is shown to align with numerical simulations. The approximation is calculated based on a noise propagation matrix obtained from the extended connectivity graph of a DQC device. In numerical simulation, we unveil the scalability enhancement in DQC for the QPUs with limited connectivity. Furthermore, we provide a simple formula to estimate the average gate fidelity, which also provides us with a heuristic method to evaluate the scalability enhancement in DQC, and a guide to optimize the structure of a DQC configuration.
\end{abstract}
\maketitle
%%                                                                  %%
%%                         Title page                               %%
%%%%%%%%%%%%%%%%%%%%%%%%%%%%%%%%%%%%%%%%%%%%%%%%%%%%%%%%%%%%%%%%%%%%%%

%%%%%%%%%%%%%%%%%%%%%%%%%%%%%%%%%%%%%%%%%%%%%%%%%%%%%%%%%%%%%%%%%%%%%%
%%                         Main text                                %%
%%                                                                  %%
  %%%%%%%%%%%%%%%%%%%%%%%%%%%%%%%%%%%%%%%%%%%%%%%%%%%%%%%%%%%%%%%%%%
  %%    Introduction
\section{Introduction}

In noisy intermediate-scale quantum (NISQ) computing, the imperfection on quantum processing units (QPUs) can not yet be corrected.
The defects on a single QPU are much more difficult to suppress as the number of qubits increases, which limits the scalability of quantum computers.
A solution for scaling up quantum computers is distributed quantum computing (DQC) \cite{CaleffiEtAlCacciapuoti2022-DQCSurvey, BarralEtAlGomez2024-RevDQC}, in which one implements global quantum circuits over multiple high-quality small-size QPUs.

The ultimate goal of DQC is to scale up quantum computing by connecting multiple QPUs.
Explicitly, one aims to implement a global unitary assisted by classical and quantum communication across different QPUs, rather than direct non-local gates.
In general, DQC can be classified into two types according to the principle of connections between two QPUs. The first type of DQC employs only local operations and classical communication (LOCC), which is called circuit knitting \cite{PendEtAlWu2020-ClusterSimDQC, MitaraiFujii2021-CircKnittingQPD, MitaraiFujii2021-CircKnitting2qGt, PiveteauSutterWoerner2022-QPD, PiveteauSutter2024-CircKnitting}.
The second type of DQC is called entanglement-assisted DQC, which employs quantum communication to share entanglement between QPUs and implement entanglement-assisted LOCC, such as state teleportation \cite{GottesmanChuang1999-UniQCmpTlpt1QbitOp} or telegating \cite{EisertEtAlPlenio2000-LclImplNonlclQGt, HuelgaEtAlPlenio2001-TlptOfUnitary, HuelgaPlenioVaccaro2002-TlptOfAngles, JiangEtAlLukin2007-DQC, MeterEtAlItoh2008-DQC, SundaramGuptaRamakrishnan2021-EffDQC, WuEtAlMurao2023-EntEffDQC, MartinezEtAlDuncan2023-DQCinQNet}.
In this paper, we consider the entanglement-assisted DQC, since the circuit knitting, has the problem of the computational overhead.

To verify the enhancement of computation power in DQC, one needs to benchmark and compare the quantum computing power of multi-QPU DQC devices and single-QPU devices.
The corresponding figures of merits include average gate fidelity (AGF)\cite{ZyczkowskiSommers2005-AvrgFdlty}, heavy output probability (HOP)\cite{CrossEtAlGambetta2019-QV}, and linear cross-entropy (LXE)\cite{GoogleEtAlMartinis2019-QSupremacy}.
The quality of the quantum computing on a quantum processor can be then quantified by averaging these figures of merits over a random sampling of quantum circuits.
In \cite{CrossEtAlGambetta2019-QV}, a randomized benchmarking based on HOP is introduced to quantify the maximum scale of a subcollection of qubits of a QPU that allows quantum advantage, which is called the quantum volume (QV) of a QPU.
The original approach of QV benchmarking is a heavy output generation test \cite{AaronsonChen2017-QSupCircSmpl} based on QV random circuits.

In this paper, we adopt different figures of merits (AGF, HOP, and LXE) to the QV random-circuit sampling, and extend it to two QPU DQC.
The extended benchmarking allows us to reveal the scalability enhancement of quantum computing in DQC.
To verify scalability enhancement in DQC, we simulate and compare the QV benchmarking for both single-QPU quantum computing and two-QPU DQC.
By definition, the average gate fidelity quantifies the faithfulness of the quantum operations implemented by a quantum device, while the heavy output probability and linear cross entropy only indicate the closeness of the measurement outputs and the ideal ones.
However, since the measurement complexity of average gate fidelity increases exponentially with the system size \cite{Nielsen2002-AGF}, it is even experimentally impossible to evaluate average gate fidelity on a quantum computer.
In contrast, the evaluation of heavy output probability and linear cross-entropy can be experimentally implemented with random circuit samplings.

\wjy{
In this paper, we show the one-to-one correspondence among average AGF $\bar{F}$, HOP $\bar{H}$, and LXE $\bar{\chi}$ in the QV random-circuit benchmarking under an approximated error model
\begin{equation}
  \bar{F}
  \;\;
  \underset{\text{QV random}}{\longleftrightarrow}
  \;\;
  \bar{H}
  \;\;
  \underset{\text{QV random}}{\longleftrightarrow}
  \;\;
  \bar{\chi}.
\end{equation}
This result allows us to estimate average gate fidelity from heavy output probability or linear cross-entropy in numerical simulation, and vice versa calculate heavy output probability and linear cross-entropy via average gate fidelity.
Employing average AGF $\bar{F}$ as the figure of merit, we derive an approximated formula of $\bar{F}$ for the QV random-circuit benchmarking of multi-QPU DQC devices under arbitrary connectivity described by extended connectivity graphs.
The theory of error model approximation is supported by numerical simulations on qiskit.
Our numerical simulations also show evidence of better average AGF of two-QPU DQC devices surpassing single-QPU devices
\begin{equation}
  \bar{F}_{DQC} \ge \bar{F}_{single}.
\end{equation}
It therefore shows the scalability enhancement of DQC under limited qubit connectivity in QPUs.}

\bigskip

The paper is constructed as follows.
In Section \ref{sec::Preliminary}, we review the concept of entanglement-assisted DQC, random-circuit benchmarking, and qubit connectivity.
In Section \ref{sec::Err_model}, we model the noises in random-circuit sampling by an approximation with single-qubit depolarizing channels, show the one-to-one correspondence among AGF, HOP, and LXE, and derive an analytical formula for average AGF.
In Section \ref{sec::num_sim_RB}, we implement numerical simulations of the QV random-circuit benchmarking and unveil the evidence for the scalability enhancement of two-QPU DQC with limited connectivity.
In the end, we discuss our results in Section \ref{sec::result} and conclude the paper in Section \ref{sec::conclusion}.

\section{Preliminary}\label{sec::Preliminary}

\subsection{Entanglement-assisted DQC}

\begin{figure}[t]
    \centering
    \includegraphics[width=0.4 \textwidth]{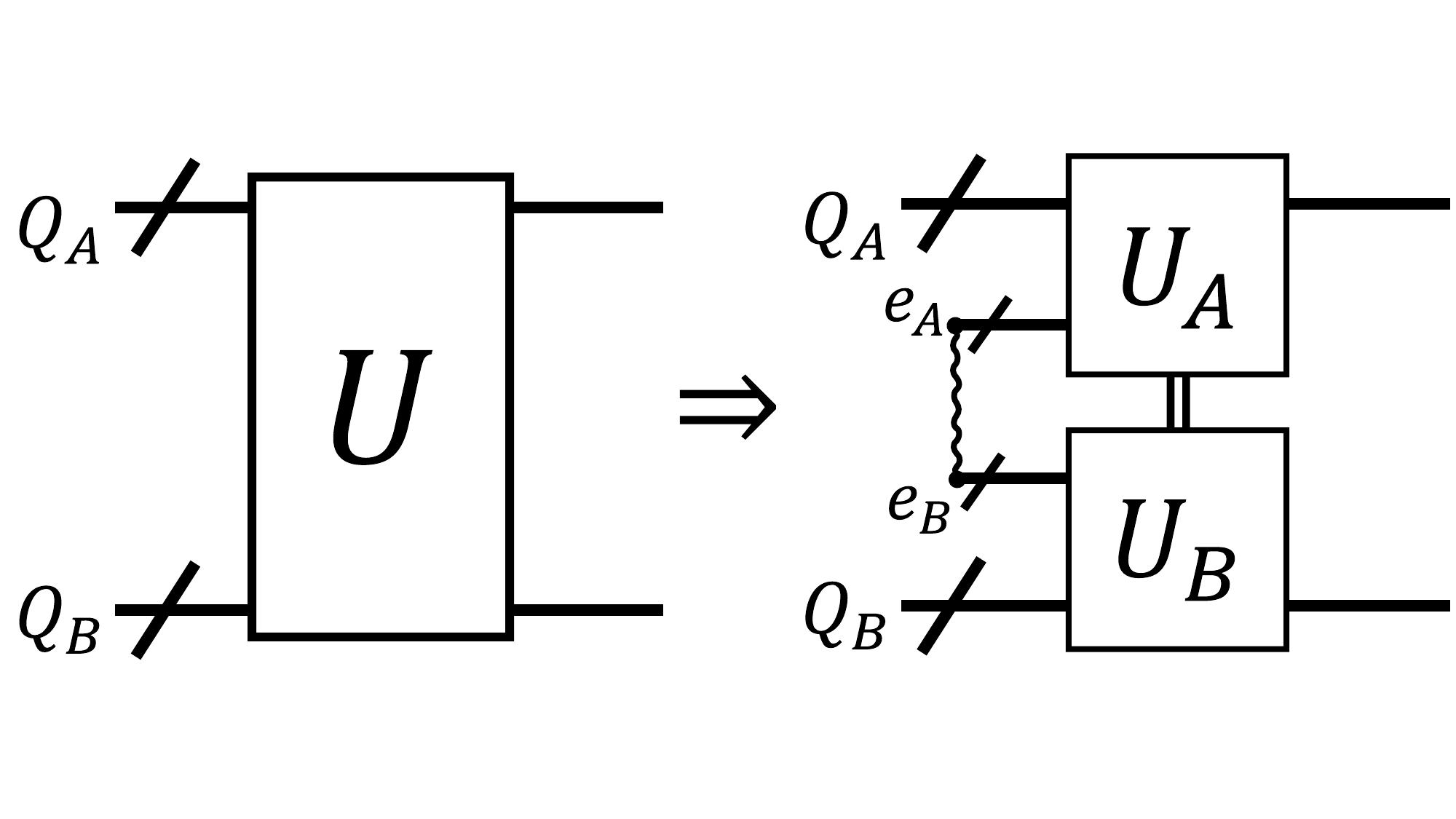}
    \caption{{Implementation of a global unitary $U$ over two QPUs $Q_A$ and $Q_B$ through local unitaries $U_A$ and $U_B$, with the help of pre-shared entangled pairs $e_{AB}$ (wiggly line) and classical communications (double line).
    % We use slash to denote multiple qubits
    }}
    \label{fig_dqc}
\end{figure}

In two-QPU entanglement-assisted DQC, one aims to execute a large quantum circuit over two small-size QPUs through local operations assisted by classical communication and pre-shared entangled pairs, i.e. entanglement-assisted LOCC, as shown in Figure~\ref{fig_dqc}.

A promising tool for achieving this is Quantum state teleportation. Quantum state teleportation is a special quantum operation that transfers a quantum state from one location to another using an entangled pair (also known as a Bell state) and local operations with classical communication (LOCC)~\cite{BennettEtAl1993-Teleportation}. Instead of directly sending the quantum state, it exploits entanglement and classical communication to transfer the state.
Expanding on this concept, there's a similar idea called gate teleportation or telegate~\cite{EisertEtAlPlenio2000-LclImplNonlclQGt, MeterEtAlItoh2008-DQC}. This proposes that the CNOT gate, a crucial quantum gate, can be executed locally using an entangled pair and LOCC. Specifically, a protocol known as the EJPP protocol (named after its creators) was introduced for this purpose~\cite{EisertEtAlPlenio2000-LclImplNonlclQGt}.

\begin{figure}[t]
    \centering
    \includegraphics[width=0.48\textwidth]{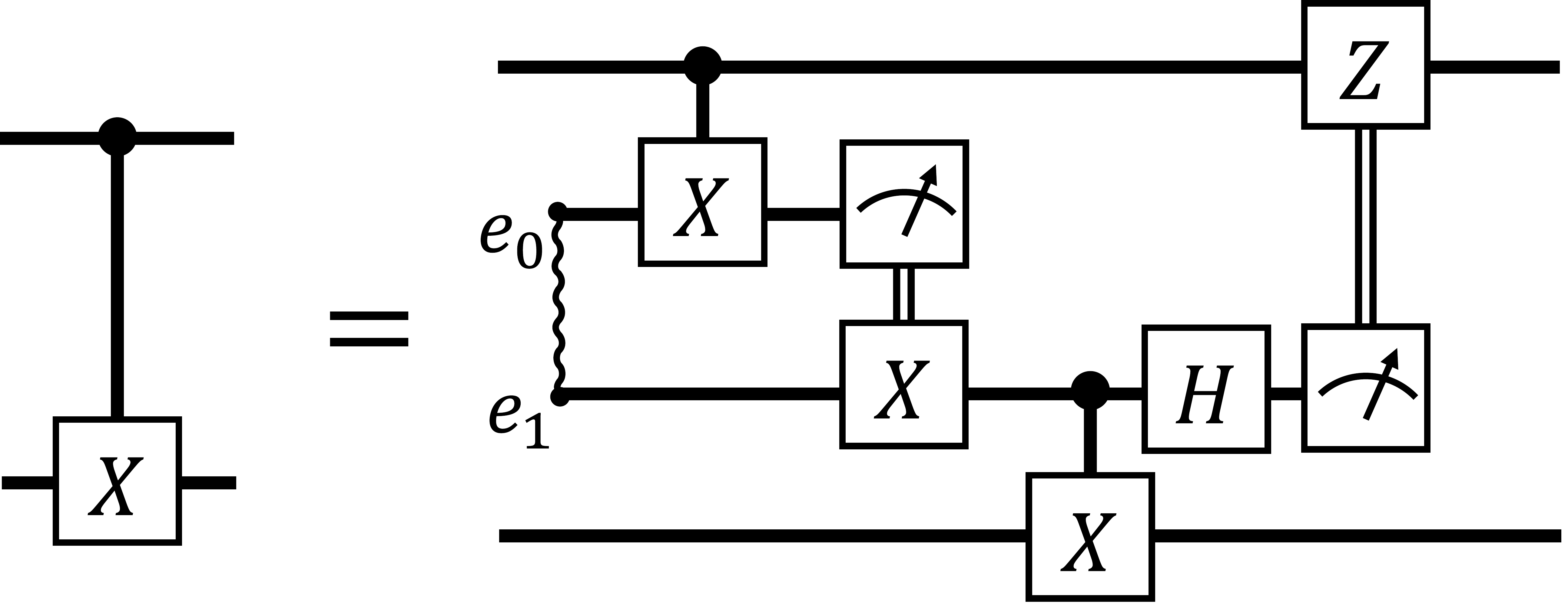}
    \caption{Implementation of a non-local CNOT gate using the EJPP protocol. The two central qubits are auxiliary qubits that share a maximally entangled pair represented by a wiggly line. Here we use the Bell state $|\Phi^+\rangle\equiv\frac{1}{\sqrt{2}}(|00\rangle+|11\rangle)$.}
    \label{fig_ejpp}
\end{figure}

The EJPP protocol's circuit, depicted in Figure~\ref{fig_ejpp}, illustrates how this telegating protocol works in practice. It enables the execution of CNOT gates across different QPUs through the clever use of entanglement and classical communication, allowing for the distributed execution of quantum circuits.

\wjy{
It is known that the CNOT gate is the only two-qubit gate in the universal set. Therefore, with the EJPP protocol, universal quantum computing can be implemented across multiple QPUs.

In our study, we aim to enhance quantum computer capabilities through DQC. We assume local QPUs have unlimited access to entangled pairs, although creating them remains challenging. Therefore, while our focus is on maximizing computational power, we also recognize the importance of minimizing entangled pair usage in circuit implementation, a topic deserving its discussion \cite{WuEtAlMurao2023-EntEffDQC}.
}

\subsection{Benchmarking via random-circuit sampling}
%In recent years, quantum computers with a limited number of imperfect qubits have emerged, necessitating systematic benchmarking methods to verify whether these devices surpass classical computers in computational power.
To validate quantum computational power, one needs a systematic benchmarking method to verify the computation on a quantum computing device.
One such benchmarking method relies on random circuit sampling, demonstrating quantum advantage by comparing outcome distributions between theoretical predictions and experimental results ~\cite{AaronsonChen2017-QSupCircSmpl}.
Ideally, one would sample circuits randomly over the Haar measure ~\cite{Mele2023-HaarMeasure}, representing a unitary invariant uniform distribution of unitary transformation.
In a practical randomized benchmarking protocol, we need random circuits that are approximately random unitaries under the Haar measure.
Prominent examples include quantum volume (QV) benchmarking\cite{CrossEtAlGambetta2019-QV} and cross-entropy (LXE) benchmarking\cite{GoogleEtAlMartinis2019-QSupremacy}.

\begin{figure}[htb]
    \centering
    \includegraphics[width=0.49 \textwidth]{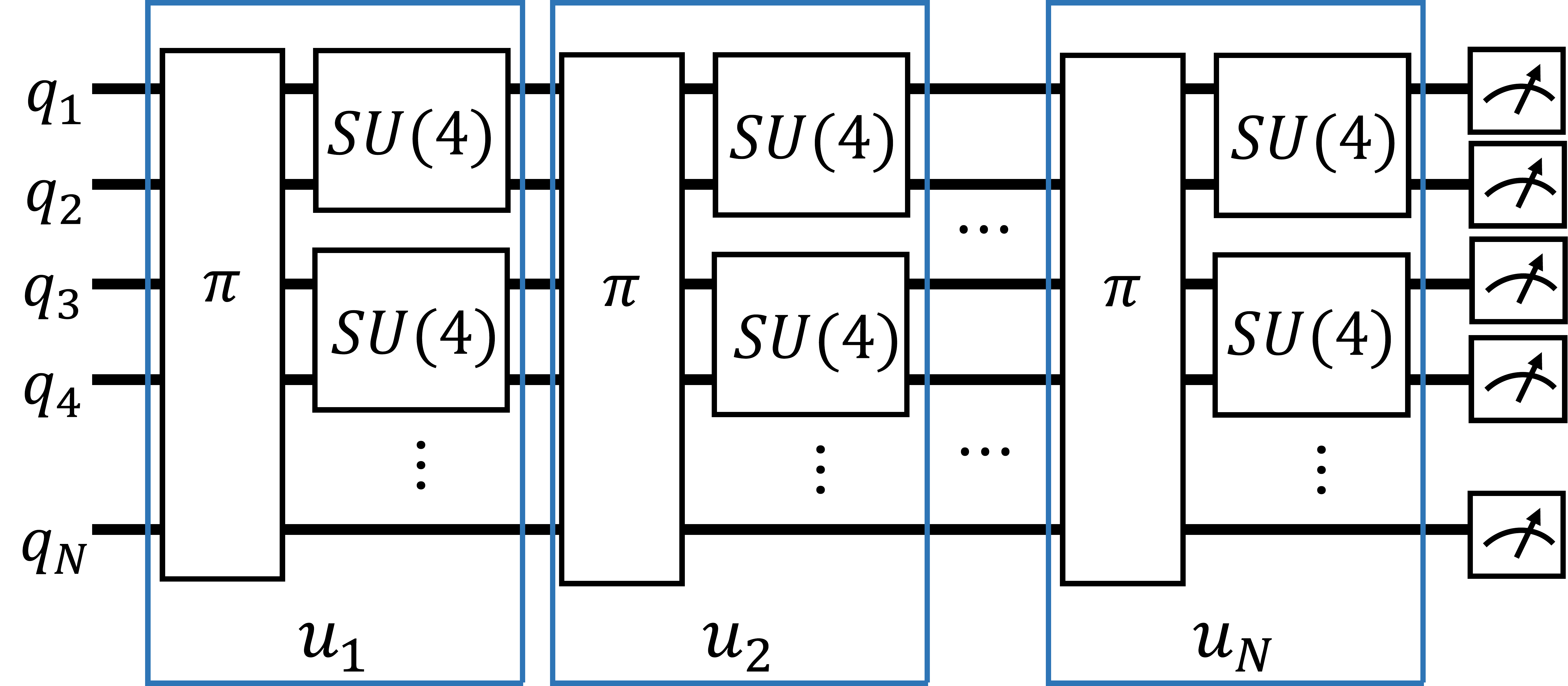}
    \caption{An N-qubit random circuit in quantum volume benchmarking. Each layer $u_i$ consists of a random permutation gate $\pi$ and several $2$-qubit Haar random $SU(4)$ gates.}
    \label{fig_qv_rc}
\end{figure}
In the QV benchmarking, which is illustrated in Fig.~\ref{fig_qv_rc}, an $N$-qubit benchmarking circuit is constituted by $N$ layers of gates. Each layer comprises a random permutation gate $\pi$ and random $2$-qubit $SU(4)$ gates sampled under the Haar measure.
To avoid unnecessary noises introduced by permutation gates, one can randomly reposition the $SU(4)$ gates instead of random permutation gates.

For a $U$ generated from a randomly sampled circuit,
\begin{equation}
  U = u_{N}.u_{N-1}\cdots u_{2}. u_{1},
\end{equation}
one calculates the theoretical distribution of outputs $q_U(x) \equiv |\braket{x|U|0}|^{2}$ on classical computers, and labels an output $x$ as heavy, if its probability $p_{x}(U)$ is greater than the median of the distribution.
The HOP $H(U)$ is then defined as the summation of the probabilities of all heavy outputs:
\begin{align}
  H(U) \equiv \sum_{x\in \mathbb{X}_U} q_x(U),
\label{def_hop}
\end{align}
where $\mathbb{X}_U\equiv\left\{x\ \text{is heavy output of}\ U\right\}$ is the set of heavy outputs, which contains $2^{N-1}$ elements.
The average HOP converges to $\frac{1+\ln{2}}{2}$ ($\approx0.85$), for the ideal devices and it is $0.5$ for the completely depolarized devices.

In the linear cross-entropy benchmarking, we calculate the linear cross-entropy $\chi_U$ between the ideal probability distribution $q_{U}(x)$ and the tested probability distribution $p_{U}(x)$ measured from the testing device, which is defined as follows,
\begin{align}
  \chi_U \equiv 2^N\sum_{x\in \{0,1\}^N}p_U(x) q_U(x) - 1.
\end{align}
%Note that both HOP and linear cross-entropy require the calculation of theoretical probability distribution, and hence only work for the devices with small numbers of qubits due to the limited computational power of classical computers.

Besides HOP and LXE, the average gate fidelity (AGF) $F(U,\Lambda)$ \cite{ZyczkowskiSommers2005-AvrgFdlty} between the target unitary $U$ and its implementation $\Lambda_{U}$ on a real device is another figure of merit for quantifying the performance of QPUs, which is defined as follows
\begin{equation}
  F(U,\Lambda_{U}) =
  \int_{\Psi} d\psi \braket{\psi|U^{\dagger} \, \Lambda_{U}(\ket{\psi}\bra{\psi})\, U|\psi},
\end{equation}
where $\Psi$ is the set of pure states under the Haar measure.
The vectorization representation of a quantum state \cite{DArianoChiribellaPerinotti2017-QInfApprch} is convenient for the calculation of AFG.
In the vectorization representation, one defines a vectorized state of a density matrix $\rho$ as follows,
\begin{align}
|\rho\rangle\rangle \equiv (\rho\otimes\mathbb{I})\sum_{i}|i\rangle\otimes|i\rangle.
\end{align}
For a pure state $\rho=|\psi\rangle\langle\psi|$, its vectorization is $|\psi\rangle\rangle \equiv |\psi\rangle\otimes|\psi\rangle^*$.
In such vectorization presentation, a unitary operator becomes
\begin{equation}
  \widehat{U}\equiv U\otimes U^*.
\end{equation}

With such vectorization the AGF of $U$ is
\begin{align}
  F(U,\Lambda_{U}) \equiv \int_{\Psi}d\psi\;\; \langle\langle\psi|\;\widehat{U}^\dag \widetilde{\Lambda}_{U}\;|\psi\rangle\rangle,
\end{align}
where $\widetilde{\Lambda}_{U}$ is the noise quantum channel of the implementation of the target unitary $U$.
Note that, in this paper, we employ the ``hat'' and ``tilde'' symbols to indicate the vectorization representation of a unitary and a quantum channel, respectively.
In randomized benchmarking, the circuits are sampled from a set of random circuits $\mathbb{U}_{RC}$, one, therefore, should further average the AGF $F(U,\Lambda_{U})$ overall the possible circuits as follows,
\begin{align}
\Bar{F} \equiv \int_{\mathbb{U}_{RC}} dU F(U,\Lambda_{U}).
\end{align}
%As you can see, although the average gate fidelity has a very straightforward meaning and good properties for analytical purposes, it is not sufficient to measure it on real devices.
We will employ these three figures of merit to evaluate the performance of DQC in the QV random-circuit sampling test.

\subsection{Connectivity of QPUs}
\begin{figure}[t]
    \centering
    \includegraphics[width=0.28 \textwidth]{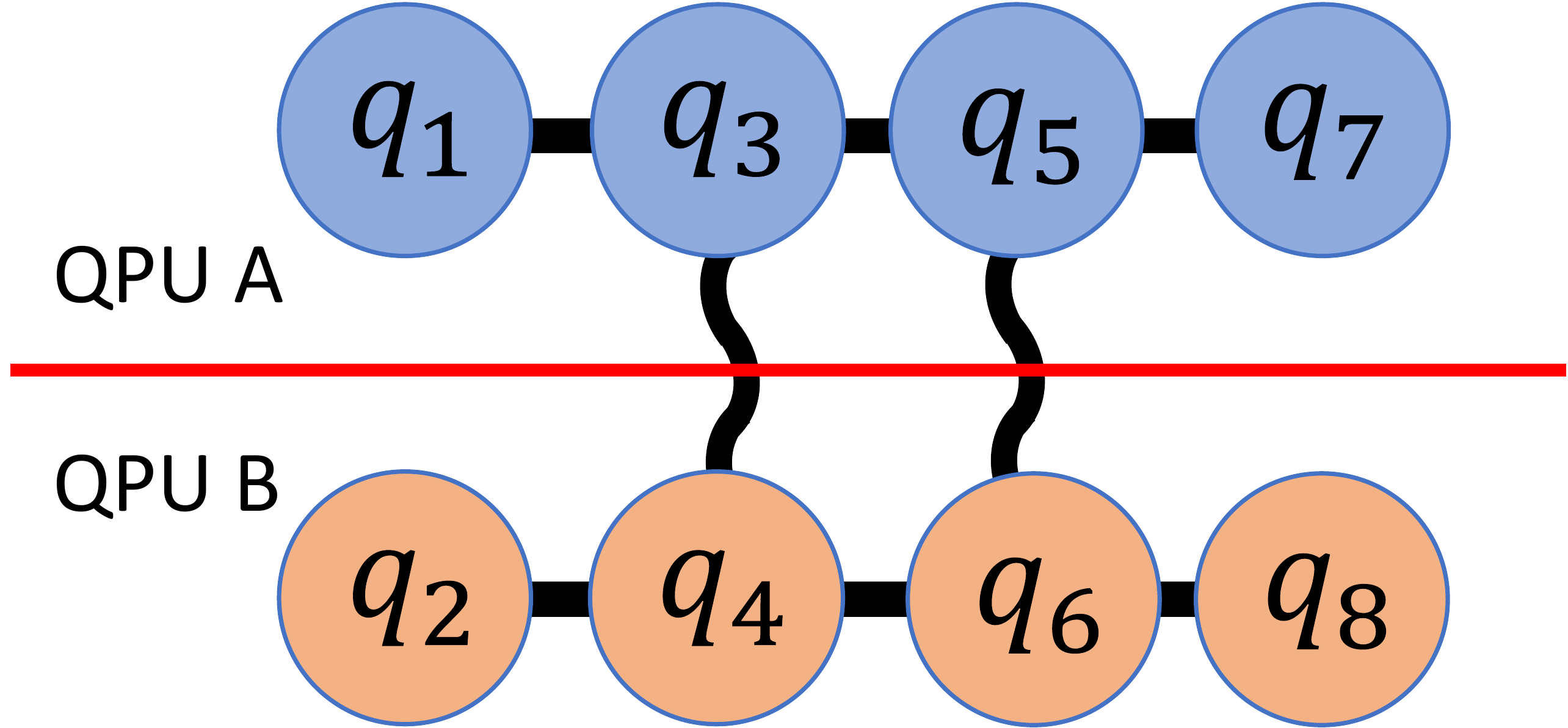}
    \caption{The extended connectivity graph of a two-QPU DQC device, which has two local QPUs with $4$ qubits on each. The solid lines denote direct physical couplings in local QPUs, while the wiggly lines stand for quantum channels across QPUs.}
    \label{fig_g-example}
\end{figure}

The connectivity constraint is a common limitation for a QPU, which prevents direct implementation of a two-qubit gate on arbitrary pairs of qubits due to a lack of direct physical coupling.
The connectivity of a QPU can be represented by a graph with its edge indicating allowed two-qubit couplings.
For a DQC device, which is composed of multiple QPUs, one can establish quantum channels sharing entanglement among qubits across the QPUs.
The topology of quantum channels in a DQC device can be also represented as a graph, in which a wiggly edge indicates a pair of shared entanglements.
We extend the connectivity graph for single-QPU devices to multi-QPU DQC devices by combining local connectivity (solid lines) and the cross-QPU quantum channels (wiggly lines) in Fig.~\ref{fig_g-example}.

\begin{figure}[t]
    \centering
    \includegraphics[width=0.4 \textwidth]{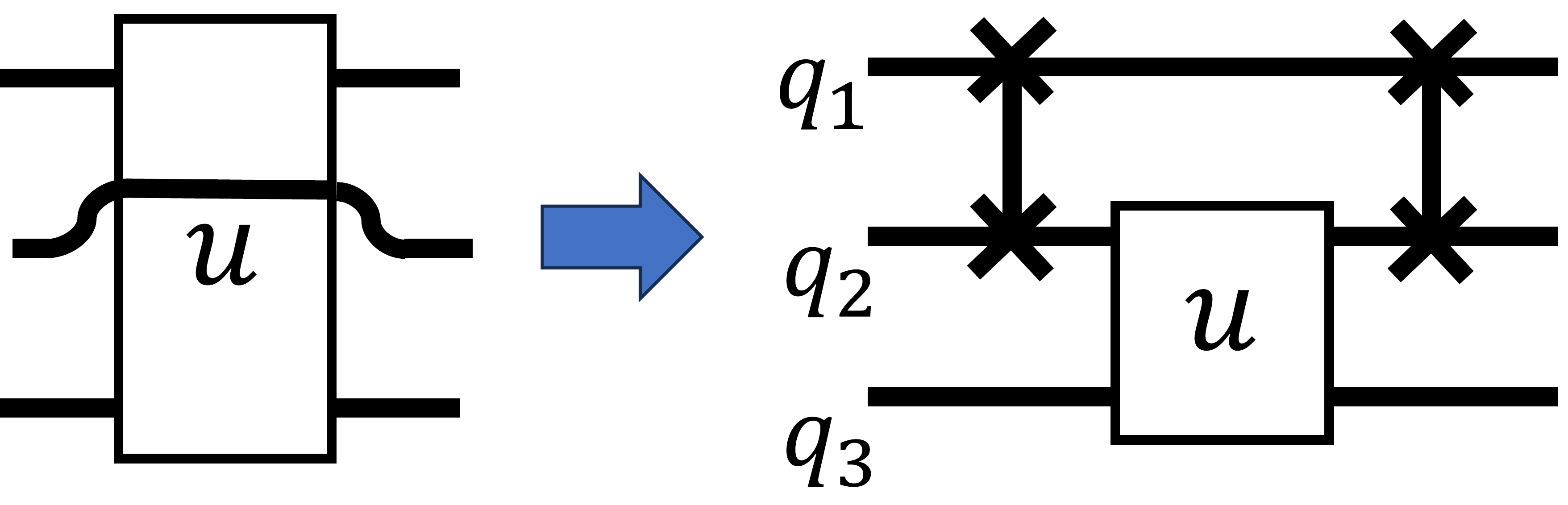}
    \caption{The implementation of a two-qubit gate on two non-connected qubits $q_{1}$ and $q_{3}$ with two additional swapping gates.
    %\gb{The figure illustrates a scenario where implementing a two-qubit gate $u$ on $Q_1$ and $Q_3$ is desired, but they lack direct connectivity, with $Q_2$ positioned between them. To address this, a series of swap operations are employed: first, a swap between $Q_1$ and $Q_2$ is performed, followed by the application of $u$ on $Q_2$ and $Q_3$, and finally, another swap between $Q_1$ and $Q_2$ is executed. This sequence effectively facilitates the desired gate operation despite the connectivity constraint.}
    }
    \label{con_swap}
\end{figure}

On a single QPU, to overcome local connectivity limitations, a common approach is to apply additional swap gates, as illustrated in Figure~\ref{con_swap}.
For high-quality performance of quantum computation under connectivity limitation, one needs to reduce the additional noises through the minimization of the swapping gates in the compilation of quantum circuits.
Such a minimization process is called qubit allocation or assignment~\cite{MaoLiuYang2023-QbtAllctnDQC}.
%However, delving into the specifics of finding this qubit allocation exceeds the scope of our current discussion.
However, even with the optimal qubit allocation, for some types of physical qubits such as superconducting qubits, the average steps of swapping among qubits in the connectivity graph of a single QPU increases very quickly as the number of qubits gets larger.

One benefit of multi-QPU DQC is that it changes and enhances the connectivity of quantum computing devices, as we can use entangled pairs shared in the quantum channels between two QPUs to reshape the connectivity graph.
The effect of such connectivity changes in DQC will be discussed in detail in the remaining sections.

\section{The error model for random-circuit benchmarking}\label{sec::Err_model}

\begin{figure*}[htb]
    \centering
    \subfloat[]{\includegraphics[width=0.98\textwidth]{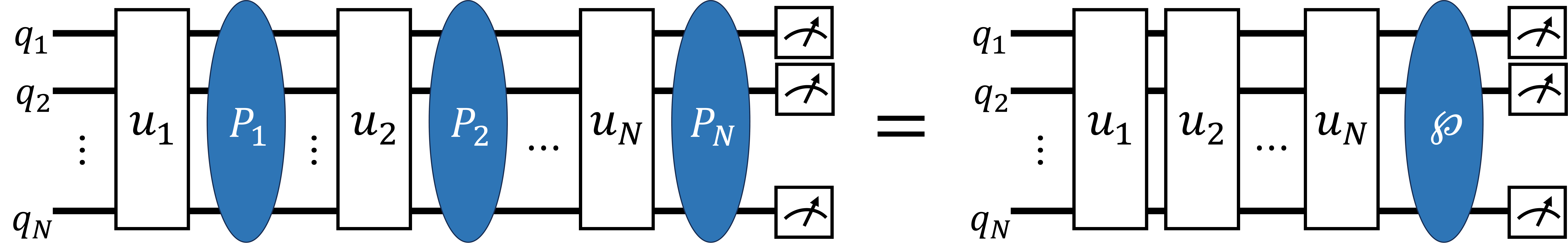}}
    \\
    \subfloat[]{\includegraphics[width=0.98\textwidth]{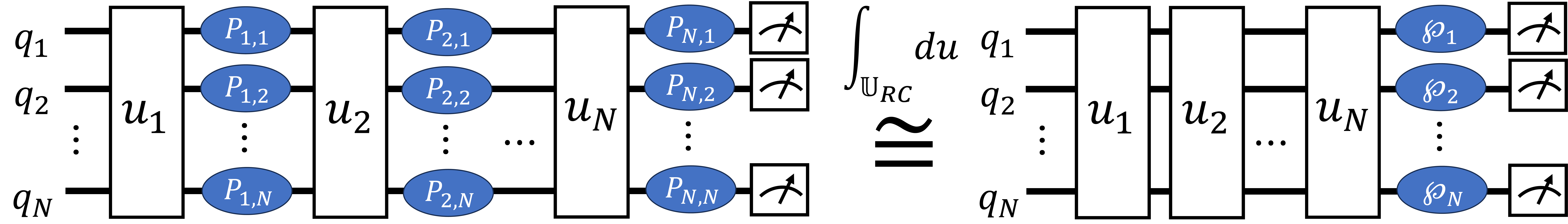}}
    \caption{(a) The global depolarizing channel (blue ellipse) comes after every layer of the random gates $u_{i}$ and acts on all the qubits globally. (b) The single-qubit depolarizing channels (blue ellipse) come after every layer of the random gates $u_{i}$ and act on each qubit individually.}
    \label{fig_error-model}
\end{figure*}

Establishing a noise model is essential for the characterization of errors in quantum devices. In the QV random-circuit sampling, the two-qubit gates are sampled from Haar random unitaries. We can therefore adopt the depolarizing channel as our fundamental noise model.
Two extreme scenarios of depolarizing noises can be employed to approximate the QV random-circuit sampling, which is visualized in Figure~\ref{fig_error-model}.
In the case of strong cross-talk between qubits, the global depolarizing noise applies, in which the depolarizing channel impacts all qubits together (see Section \ref{sec::global_depolarizing}).
In the other situation where cross-talk among qubits is negligible, the single-qubit depolarizing noise is applicable (see Section \ref{three_two}).
For the single-qubit depolarizing noise, the depolarizing channel affects each qubit individually.
%In the QV benchmarking, where there is a random permutation of the qubits which will let these two approximations become equivalent, as we will see in the latter section.
In the QV benchmarking, which necessitates the random permutation of qubits, these two approximations are shown to be equivalent in Section~\ref{three_two}.

With these error models, one can evaluate the benchmarking of QV random-circuit sampling employing either the average gate fidelity, heavy output probability, or linear cross-entropy as the figure of merit.
Under the single-qubit error model, we show the one-to-one correspondence of these three metrics in the QV random-circuit sampling.
Such correspondence enables us to treat heavy output probability and linear cross-entropy as the function of the average gate fidelity, which benefits us in two folds.
Namely, it provides us with an analytical tool for studying HOP and LXE through the calculation of AGF, as well as the potential to estimate the AVG from experimental data of HOP or LXE.

\subsection{Global depolarizing noise}
\label{sec::global_depolarizing}
In the vectorization representation, the effect of the N-qubits depolarizing channel acting on N-qubits state $|\rho\rangle\rangle$ is
\begin{align}
\label{Nq_depolarising}
  \widetilde{D}_P^{(N)}|\rho\rangle\rangle = P|\rho\rangle\rangle+\frac{(1-P)}{2^N}|\mathbb{I}_{2^N}\rangle\rangle,
\end{align}
where we call $P$ the preserving factor of the depolarizing channel.
For $P=1$, one has a perfect identity operator, while for $P=0$, one obtains a complete depolarizing channel.
As shown in Fig. \ref{fig_error-model} (a), the quantum circuit implementation of $U = u_{N}.\cdots.u_{1}$ under the global depolarizing noise is then given as follows,
%\begin{align}
%\Tilde{U}(\rho) &= (\Tilde{u}_N\circ\dots\circ\Tilde{u}_2\circ\Tilde{u}_1)(\rho)\notag \\
%&= (\Tilde{D}_{P_N}^{all}\circ\hat{u}_N\circ\dots\circ\Tilde{D}_{P_2}^{all}\circ\hat{u}_2\circ\Tilde{D}_{P_1}^{all}\circ\hat{u}_1)(\rho)\label{gea1}\\
%\Tilde{D}_{P}^{all}(\rho)&\equiv P\rho + (1-P)\frac{\mathbb{I}}{2^N}\;\;\;with\;P\in [0,1]\notag
%\end{align}
%where N is the number of qubits and we use $\hat{u}$ to denote the unitary channel, i.e. $\hat{u}(\rho)=u\rho u^\dag$,
\begin{align}
  \widetilde{U} &= \widetilde{u}_N.\cdots.\widetilde{u}_2.\widetilde{u}_1
  \nonumber\\
  & =\widetilde{D}_{P_N}^{(N)}.\hat{u}_N.\cdots.\widetilde{D}_{P_2}^{(N)}.\hat{u}_2.\widetilde{D}_{P_1}^{(N)}.\hat{u}_1.
  \label{gea1}
\end{align}
Since the global depolarizing channel commutes with all other channels, we can simplify Eq.~(\ref{gea1}) by moving all depolarizing channels to the end of the circuit and retaining only one depolarizing channel with an effective global preserving factor $\wp$,
\begin{equation}
  \widetilde{U} = \widetilde{D}_{\wp}^{(N)}. \hat{U}, \;\;\text{with } \wp =\prod_j P_j.
\end{equation}
Under this global depolarizing noise approximation, the average gate fidelity for a given random circuit $U$ can be then expressed as a function of the effective preserving factor,
%F_U(P) &= \int_{\Psi}d\psi\;\langle\psi| U^\dag\widetilde{D}_{P}^{all}(U|\psi\rangle\langle\psi| U^\dag)U|\psi\rangle\notag\\
%&=P + \frac{1-P}{2^N}\label{gea_gf}
\begin{align}
  F_U(\wp)&= \int_{\Psi}d\psi\;\; \langle\langle\psi|\widehat{U}^\dag\widetilde{D}_{\wp}^{(N)}\widehat{U}|\psi\rangle\rangle %  = \wp + \frac{1-\wp}{2^N}.
  \nonumber\\
  & = \wp + \frac{1}{2^N}(1-\wp).
\label{gea_gf}
\end{align}
Similarly, we can determine the HOP as a function of the preserving factor:
%H_U &= \sum_{x\in X_U} \langle x|\widetilde{D}_{P}^{all}(U|0^N\rangle\langle 0^N| U^\dag)| x\rangle
%\notag\\
%&=H_U^{ideal} P + \frac{1-P}{2}
\begin{align}
  H_U(\wp) & = \sum_{x\in X_U}\langle\langle x|\widetilde{D}_{\wp}^{(N)}\widehat{U}|0^N\rangle\rangle
  \nonumber\\
  & = H_U^{ideal} \wp + \frac{1}{2}(1-\wp).
\label{gea_hop}
\end{align}
Here, $H_U^{\text{ideal}}$ is the ideal HOP for the circuit $U$, and the $X_U$  is the set of heavy outputs of $U$.
Regarding the linear cross-entropy, we can write it as
%\chi_U \equiv&2^N\sum_{x} \langle x|[P\;U|0^N \rangle\langle0^N|U^\dag + (1-P)\frac{\mathbb{I}_{2^N}}{2^N}]|x\rangle |\langle x|
%U|0^N \rangle|^2 - 1 = P\chi_U^{ideal}\label{gea_xeb}
\begin{align}
  \chi_U
  & = 2^N\sum_{x} \langle\langle x|\widetilde{D}_{\wp}^{(N)}\widehat{U}|0^N\rangle\rangle \langle\langle x|\widehat{U}|0^N\rangle\rangle -1
  \nonumber\\
  &=\chi_U^{ideal} \; \wp.
\label{gea_xeb}
\end{align}
Although this approximation seems rough, later analysis reveals its effectiveness in illuminating the relationship between average gate fidelity, HOP, and linear cross-entropy.

\subsection{Single-qubit depolarizing noise}\label{three_two}

In the single-qubit error approximation, we apply two single-qubit depolarizing channels after each two-qubit gate. As shown in Fig. \ref{fig_error-model} (b), in the vectorization presentation, the quantum channel of a random circuit under single-qubit depolarizing noises can be written as
%\begin{align}
%\widetilde{U}(\rho) &= (\widetilde{D}_{\vec{P}_N}\circ\widehat{u}_N\circ\dots\circ\widetilde{D}_{\vec{P}_2}\circ\widehat{u}_2\circ\widetilde{D}_{\vec{P}_1}\circ\widehat{u}_1)(\rho)\label{sqea1}\\
%\widetilde{D}_{\vec{P}}(\rho)&\equiv (\bigotimes_ {q\in \mathbb{Q}}\widetilde{D}^{(q)}_{\vec{P}_q})(\rho)\notag
%\end{align}
\begin{align}
\label{sqea1}
  \widetilde{U} &= \widetilde{D}_{\vec{P}_N}.\widehat{u}_N
  \,.\,\cdots\,.\,\widetilde{D}_{\vec{P}_2}.\widehat{u}_2
  \,.\,\widetilde{D}_{\vec{P}_1}.\widehat{u}_1,
\end{align}
where $\widetilde{D}_{\vec{P}_{i}}$ is the tensor product of the single qubit depolarizing channels and $\vec{P}_{i}=(P_{i,q_1},P_{i,q_2},...,P_{i,q_N})$ is a vector, whose $q$ component $P_q$ is the preserving factor of the single-qubit deodorizing channel acting on the qubit $q$,
\begin{equation}
  \widetilde{D}_{\vec{P}}\equiv \bigotimes_{q\in \mathbb{Q}}\widetilde{D}_{P_q}^{(1)}.
\end{equation}
Here, $\mathbb{Q}=\{q_{1},q_{2},...,q_{N}\}$ is the set of qubits of a QPU. %Note that for an odd number of qubits, there will be one qubit in each layer without a gate acting on it.

Unlike the global depolarizing channel, the single-qubit depolarizing channel typically does not commute with other channels.
However, in the present scenario, our unitary operators are sampled randomly from the Haar uniform distribution.
Such a random sampling allows us to make an assumption that $\widetilde{D}_{\vec{P}}\,.\,\widehat{u}=\widehat{u}\,.\,\widetilde{D}_{\vec{P}}$ holds for all unitary channels. %We will justify this assumption later with results \gb{derived from} numerical simulations.
%We made this assumption based on the fact that the random SU(4) gate under the Haar \gb{uniform} distribution commutes with the single-qubit depolarizing channel in average, i.e.
We arrived at this assumption by considering the behavior of a random SU(4) gate generated according to the Haar uniform distribution in relation to the single-qubit depolarizing channel. Specifically, we observe that, on average, the random SU(4) gate commutes with the single-qubit depolarizing channel. Mathematically, this can be expressed as
\begin{align}
\label{eq::1q_depol_commute}
\int_{U\in SU(4)} dU\;\;[\widehat{U}, \widehat{\mathbb{I}}\otimes \widetilde{D}_{P_{q}}^{(1)}] = 0.%, \;\;\forall P\in[0,1].
\end{align}
A detailed proof of this commutation relation is provided in Appendix \ref{app_comm_hy}.
In this single-qubit depolarizing-noise model, our channel only contains two-qubit gates and single-qubit depolarising channels.
We can therefore rearrange the order of operations by moving all depolarizing channels to the end of the circuit.
\begin{equation}
  \widetilde{U} =
  \widetilde{D}_{\vec{\wp}}\, .\,\widehat{U}
\end{equation}
Note that the effects of noises are independent of the circuit.
%For all quantum states $\rho$, we now define an associated probability vector $\mathcal{P}^\rho$, where $\mathcal{P}^\rho_n\equiv \langle n|\rho|n\rangle$.
The vector of the measurement probability distribution $\boldvec{p}(U) \equiv \{p_{U}(n)\}_{n}$ under the single-qubit depolarizing noise model is then given by
\begin{equation}
  p_{n}(U) \equiv
  \langle\langle n| \widetilde{D}_{\vec{\wp}}.\widehat{U}|0,...,0\rangle\rangle.
\end{equation}
Suppose $\boldvec{q}(U)\equiv \{p_{n}(U)\}_{n}$ be the ideal distribution
\begin{equation}
  q_{n}(U) \equiv
  \langle\langle n| \widehat{U}|0,...,0\rangle\rangle..
\end{equation}
One can obtain $\boldvec{p}_{U}$ from a Markov chain $\mathcal{D}_{\vec{\wp}}$ acting on the ideal outcome distribution $\boldvec{q}_{U}$,
\begin{equation}\label{sqea2}
  \boldvec{p}(U)
  = \sum_{m\in\{0,1\}^{\otimes N}}\mathcal{D}_{n|m}(\vec{\wp})q_{m}(U)
  = \mathcal{D}(\vec{\wp}).\boldvec{q}(U),
\end{equation}
where $\mathcal{D}(\vec{\wp}) \equiv \{\mathcal{D}_{n|m}(\vec{\wp})\}_{n,m}$ and $\mathcal{D}_{n|m}(\vec{\wp}) \equiv \langle\langle n|\widetilde{D}_{\vec{\wp}}|m\rangle\rangle$.
The Markov matrix describes the effects of the effective single-qubit depolarizing channels $\widetilde{D}_{\wp_{i}}^{(1)}$ acting at the end of the circuit shown in Fig. \ref{fig_error-model} (b), which can be formulated as a tensor product of single-qubit Markov matrix,
\begin{align}
\mathcal{D}(\vec{\wp})
&=  \bigotimes_{q\in \mathbb{Q}} \left(\wp_q\begin{bmatrix}
1 & 0 \\
0 & 1
\end{bmatrix} + \frac{1-\wp_q}{2}\begin{bmatrix}
1 & 1 \\
1 & 1
\end{bmatrix}\right).
\end{align}
Due to the commutativity in Eq. \eqref{eq::1q_depol_commute}, the average gate fidelity is the diagonal element of the Markov matrix,
%\begin{comment}
%\begin{align}
%F_U(\vec{P}) &= \int_{\Psi}d\psi\;\langle\psi| U^\dag\widetilde{D}_{\vec{P}}(U|\psi\rangle\langle\psi| U^\dag)U|\psi\rangle\notag\\
%&= \langle 0^N|\widetilde{D}_{\vec{P}}(|0^N\rangle\langle 0^N|)|0^N\rangle = \prod_{q\in \mathbb{Q}}\frac{1+\vec{P}_q}{2}\label{gf_markov}
%\end{align}
%\end{comment}
\begin{align}\label{gf_markov}
  F_U(\vec{\wp}) &=
  \int_{\Psi}d\psi\;\; \langle\langle\psi|\widehat{U}^\dag\widetilde{D}_{\vec{\wp}}\widehat{U}|\psi\rangle\rangle
  \nonumber\\
  & = \langle\langle 0,...,0|\widetilde{D}_{\vec{\wp}}|0...,0\rangle\rangle.
\end{align}
It implies that the average gate fidelity is independent of $U$.
We hence obtain the average AGF, $\bar{F} = F_{U}(\vec{P})$,
\begin{equation}
\label{eq::agf_eff_p}
  \bar{F} = \prod_{q\in \mathbb{Q}}\frac{1+\wp_q}{2},
\end{equation}
which can be also expressed by an effective preserving factor $\bar{\wp}$ of a global depolarizing channel
\begin{equation}
\label{eq::avg_AGF_U_indep}
  \bar{F} = \bar{\wp} + \frac{1}{2^N}(1-\bar{\wp})\;\;
  \text{ with }
  \bar{\wp} \equiv \frac{\prod_{q\in\mathbb{Q}}(1+\wp_{q})-1}{2^{N}-1}.
\end{equation}
The determination of the effective preserving factor $\vec{\wp}$ of single-qubit depolarizing channels will be discussed in the next section.

\bigskip
To calculate the heavy output probability, we first express it as the inner product between the probability vector $\boldvec{p}(U)$ and the heavy output vector $\boldvec{h}(U)$,
%\begin{comment}
%\begin{align}
%H^{ideal} &= \sum_{x\in X_U}\mathcal{P}^{\widehat{U}(|0^N\rangle\langle 0^N|)}_x = \sum_{n\in X} h_{U}^n\mathcal{P}^{\widehat{U}(|0^N\rangle\langle 0^N|)}_n\\
%h_{U}^n &\equiv
%\begin{cases}
%1\;\;if\;n \in X_U\\
%0\;\;if\;n \notin X_U
%\end{cases}
%\end{align}
%\end{comment}
\begin{equation}
  H_{ideal}(U) = \boldvec{h}(U)^{T}.\boldvec{p}(U),
\end{equation}
where
\begin{align}
%H_{ideal}(U) &=
%\boldvec{h}(U).\boldvec{p}(U)
%\sum_{n\in\{0,1\}^N}h_n(U) \,p_{n}(U)\\
h_n(U) &=
\begin{cases}
1\;\;,\;n \text{ is heavy}\\
0\;\;,\;n \text{ is not heavy}
\end{cases}.
\end{align}
For each $U$, one can reorder the probability distribution $\boldvec{p}$ in a descending order through a permutation $\Pi_{U}$, such that $p_{\Pi(0)}\ge p_{\Pi(1)}\ge\cdots\ge\ p_{\Pi(2^{N})}$. The ideal HOP for $U$ is therefore the sum of the probabilities of the top half of the probability $\Pi_{U}.\boldvec{p}(U)$ filtered by a canonical heavy output vector $\bar{\boldvec{h}}$,
\begin{equation}
  H_{\text{ideal}}(U) = \bar{\boldvec{h}}^{T}.\Pi_{U}.\boldvec{p}(U),
\end{equation}
where
\begin{align}
  \bar{h}_n&=
  \begin{cases}
  1\;\;if\;n < 2^{N-1}\\
  0\;\;if\;n \geq 2^{N-1}
  \end{cases}.
\end{align}
The ideal average HOP $\bar{H}_{\text{ideal}}$ can be then obtained from an average probability vector $\bar{\boldvec{p}}$ filtered by $\bar{\boldvec{h}}$,
\begin{equation}
  \bar{H}_{\text{ideal}} = \bar{\boldvec{h}}^{T}.\bar{\boldvec{p}},
\end{equation}
where
\begin{align}
  \bar{\boldvec{p}} = \int_{U\in \mathbb{U}_{RC}} dU\; \Pi_U. \boldvec{p}(U).
\end{align}
Under the QV random-circuit sampling, we can incorporate errors into the average HOP by inserting a symmetrized Markove matrix $\bar{\mathcal{D}}(\vec{\wp})$
\begin{align}
\label{hop_markov}
  \Bar{H}(\vec{\wp}) =
  \Bar{\boldvec{h}}^T.\Bar{\mathcal{D}}(\vec{\wp}) \,.\,\Bar{\boldvec{p}},
\end{align}
where $\bar{\mathcal{D}}(\vec{\wp})$ is symmetrized under all permutations $\Pi$ of the outputs $\{0,...,2^{N}-1\}$ in the $S_{2^{N}}$ group,
\begin{equation}
  \Bar{\mathcal{D}}(\vec{\wp}) \equiv \frac{1}{2^{N}!}\sum_{\Pi \in \mathbf{S}_{2^N}}\Pi^T \mathcal{D}(\vec{\wp})\Pi.
\end{equation}
After averaging over all permutation, $\Bar{\mathcal{D}}(\vec{\wp})$ is determined as
\begin{align}\label{avg_markov}
  \Bar{\mathcal{D}}(\vec{\wp}) &=
  \begin{bmatrix}
    \Bar{F} & \frac{1-\Bar{F}}{2^N-1} & \dots  & \frac{1-\Bar{F}}{2^N-1} \\
    \frac{1-\Bar{F}}{2^N-1} & \Bar{F}  & \dots  & \frac{1-\Bar{F}}{2^N-1} \\
    \vdots & \vdots  & \ddots & \vdots \\
    \frac{1-\Bar{F}}{2^N-1} & \frac{1-\Bar{F}}{2^N-1} & \dots  & \Bar{F}
  \end{bmatrix}.
\end{align}
It becomes an effective global depolarizing channel with the effective preserving factor givin in Eq. \eqref{eq::avg_AGF_U_indep}.
A detailed derivation of Eq. \eqref{hop_markov} is provided in Appendix \ref{sec::proof_effective_markov_in_HOP}.

Consequently, we arrive at the average HOP $\bar{H}$ as a functions of the average AGF $\bar{F}$, which are them given by the effective global preserving factor $\bar{\wp}$,
\begin{align}
\label{hop_gf}
  \bar{H}(\vec{\wp}) &= \bar{H}_{ideal}\frac{2^N\Bar{F}-1}{2^N-1}+(1-\Bar{F})\frac{2^{N-1}}{2^N-1}
  \nonumber\\
  & = \bar{H}_{ideal}\;\bar{\wp}+\frac{1}{2}(1-\bar{\wp}).
\end{align}
The same result holds for the LXE
\begin{align}\label{xeb-gf}
  \bar{\chi}(\vec{\wp})&= \frac{2^N\Bar{F}-1}{2^N-1}\bar{\chi}_{ideal}
  = \bar{\chi}_{ideal} \; \bar{\wp}.
\end{align}
A linear relation between linear cross-entropy benchmarking and heavy output testing can be then derived,
\begin{align}
\bar{H}(\vec{\wp})= \frac{1}{2}+\frac{1}{2}\ln 2 \frac{\bar{\chi}(\vec{\wp})}{\bar{\chi}_{ideal}}.
\label{hop-xeb}
\end{align}
These two equations imply the one-to-one correspondence among average HOP, LXE, and average AGF.
This one-to-one correspondence is exactly the same as their relation in the global depolarizing noise model.
It implies the equivalence between global depolarizing noise and single-qubit depolarizing noise in the random-circuit benchmarking.
In the large size limit ($2^N \gg 1$), this one-to-one correspondence can be simplified to
\begin{align}
\label{eq::one-to-one_simple}
\bar{H}(\vec{\wp})
%&\approx\braket{H}_{ideal}\Bar{F} + \frac{1-\Bar{F}}{2}
\approx \frac{1+\Bar{F}\ln{2}}{2},
\;\;\;\;
\bar{\chi}(\vec{\wp})&\approx \Bar{F} \;\bar{\chi}_{ideal}.
\end{align}
%\begin{comment}
%We can also use \gb{Eq.~\ref{avg_markov}} %at
%\gb{to calculate} the linear cross-entropy,  %benchmarking,
%as shown in Eq.(\ref{xeb-gf}).
%\begin{align}
%\braket{\chi(\vec{P})}&= 2^N \sum_{n,m\in X}\Bar{\mathcal{P}}^n\Bar{D}(\vec{P})^m_n\Bar{\mathcal{P}}_m - 1=\frac{2^N\Bar{F}-1}{2^N-1}\braket{\chi}^{ideal}\label{xeb-gf}
%\end{align}
%\end{comment}
%At the large size limit, we have $\frac{\braket{\chi(\vec{P})}}{\braket{\chi}^{ideal}}\approx \Bar{F}$.
In the quantum volume benchmarking, if devices pass the test, then we must have $\bar{H}(\vec{\wp})>\frac{2}{3}$. So the threshold for the linear cross-entropy, by the suggestion of the quantum volume benchmarking, is $\bar{\chi}(\vec{\wp})>\frac{1}{3\ln{2}}\bar{\chi}_{ideal}$, and similar for the average gate fidelity. Another important consequence of Eq.~(\ref{hop_gf}) and Eq.~(\ref{xeb-gf}) is that now HOP and LXE are the linear functions of the average gate fidelity.
For arbitrary two QPUs with the effective single-qubit preserving factors $\vec{\wp}_{\mathbb{Q}}$ and $\vec{\wp}_{\mathbb{Q}'}$, the linearity implies the following monotonic relation,
\begin{align}
\label{monotonic}
  \Bar{F}(\vec{\wp}_{\mathbb{Q}})\geq\Bar{F}(\vec{\wp}_{\mathbb{Q}'})
  & \Leftrightarrow \Bar{H}(\vec{\wp}_{\mathbb{Q}})\geq\Bar{H}(\vec{\wp}_{\mathbb{Q}'})\nonumber\\
  & \Leftrightarrow \bar{\chi}(\vec{\wp}_{\mathbb{Q}})\geq\bar{\chi}(\vec{\wp}_{\mathbb{Q}'}).
\end{align}
As a result, we only need to employ average gate fidelity to evaluate the performance of QPUs, which is our main focus in the next section.

\subsection{The error model for DQC under connectivity constraints}

In the previous section, we derive a one-to-one correspondence relation among different metrics QV random-circuit benchmarking, which allows us to employ average AGF to evaluate quantum computing devices.
According Eq. \eqref{gf_markov} and \eqref{eq::avg_AGF_U_indep}, the average AGF is determined by the effective preserving factors $\vec{\wp} = (\wp_{q_{1}},...,\wp_{q_{N}})$.
In this section, we will establish a theory for determining the effective vector $\vec{\wp}$ under connectivity constraints.

\begin{figure*}[htb]
    \centering
    \hfill
    \subfloat[]{\includegraphics[width=0.2\textwidth]{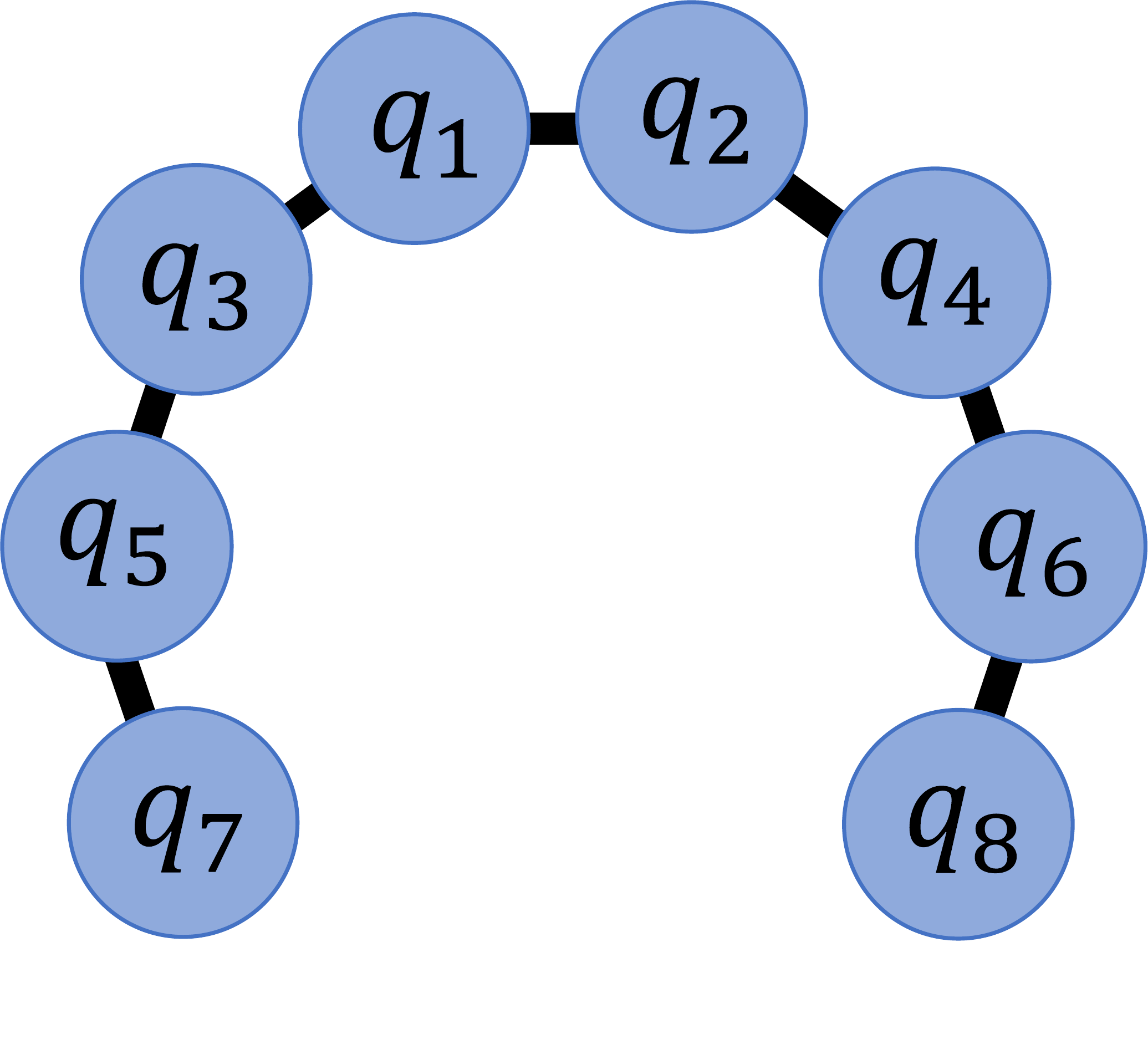}}
    \hfill
    \subfloat[]{\includegraphics[width=0.2\textwidth]{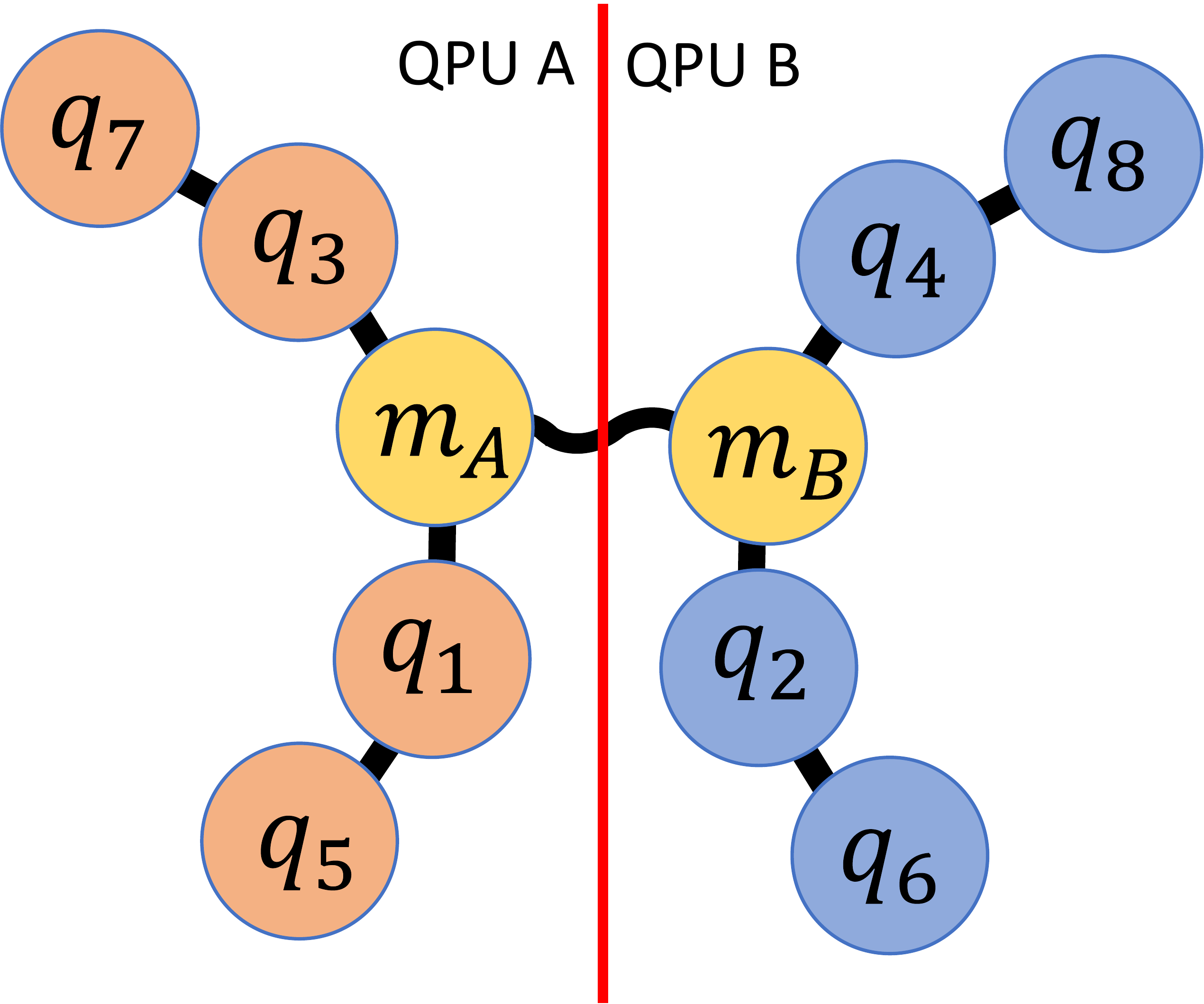}}
    \hfill
    \subfloat[]{\includegraphics[width=0.2\textwidth]{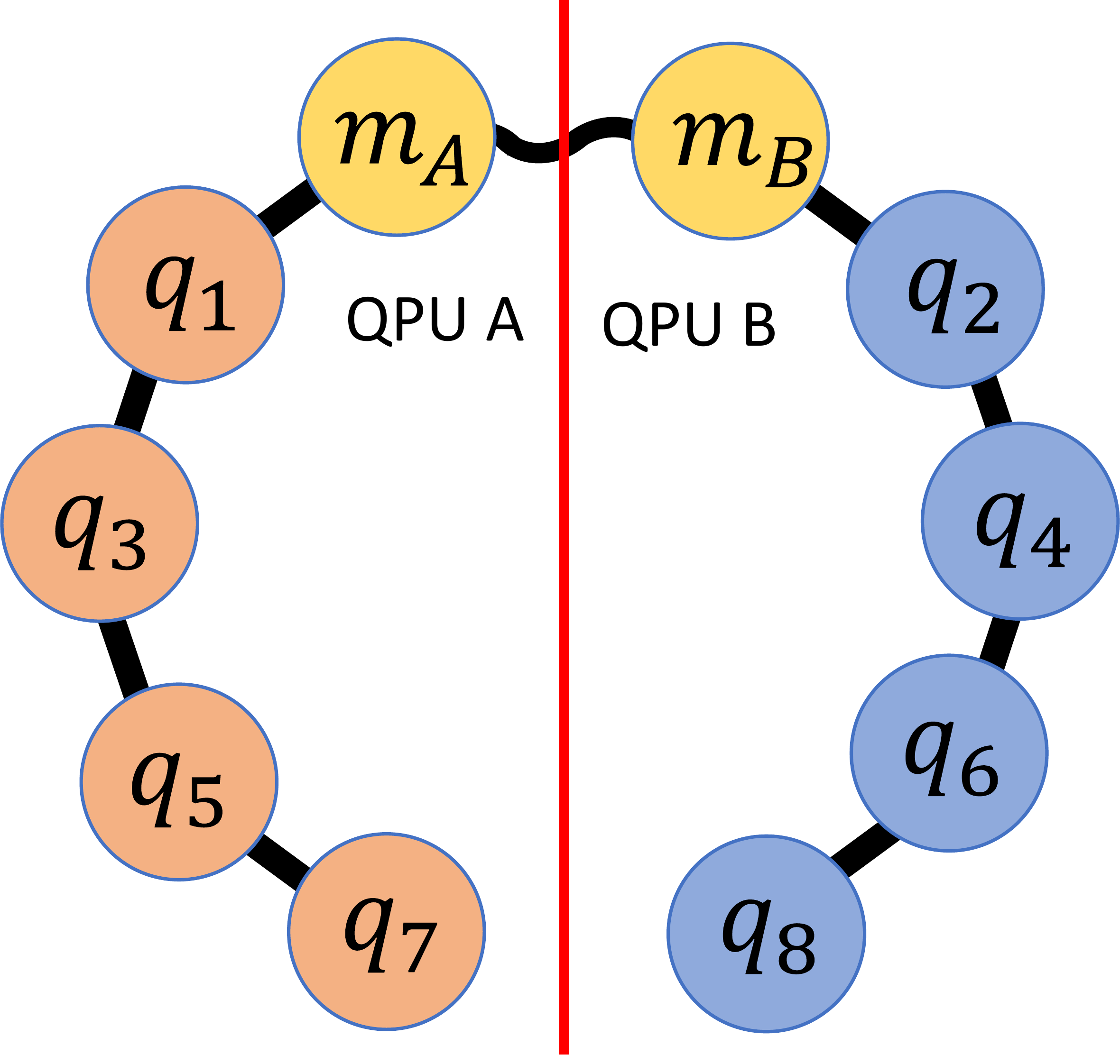}}
    \hfill{ }
    \caption{%The (a) and (b) are the same graph we used in Figure~\ref{fig_g_table}. The (c) also so had a 1D graph local QPU, the only difference between (b) and (c) is the location of the auxiliary memory qubit. As you can see, the graph of (a) and (c) is the same if one replaces the auxiliary memory qubits with a coupling between $q_0$ and $q_1$.
    Extended connectivity graph. The distinction between graphs (b) and (c) is the positioning of the auxiliary memory qubits. Notably, the graphs (a) and (c) coincide if the auxiliary memory qubits are replaced with a coupling between $q_1$ and $q_2$.
    (a) A 1D connectivity graph of a single QPU with $8$ qubits.
    (b) A two-1D-QPU DQC connectivity graph with $8$ working qubits in total. The memory qubit on each QPU stores an entangled pair between the two QPUs.
    (c) A two-1D-QPU DQC connectivity graph resulting a new extended 1D connectivity graph.}
    \label{fig_3d1-graph-table}
\end{figure*}

\bigskip

We start our analysis with the fully connected single-QPU device, which is the simplest case. An $N$-qubit quantum volume circuit has N layers of gates.
In each layer, there are $\lfloor\frac{N}{2}\rfloor$ SU(4) gates acting on N qubits.
In the single-qubit depolarizing noise model, each SU(4) gate is followed by two single-qubit depolarizing channels. There are in total $2N\lfloor\frac{N}{2}\rfloor$ depolarizing channels acting on the $N$ qubits.
Here, we assume the preserving factor $P_{q}$ of a single-qubit depolarizing channel is a constant for a qubit $q$, independent of layers.
In average, there are $2\lfloor\frac{N}{2}\rfloor$ effective single-qubit depolarizing channels in the end of the circuit.
The effective preserving factor is
\begin{equation}
  \wp_{q} = P_{q}^{2\lfloor\frac{N}{2}\rfloor}.
\end{equation}
According to Eq. \eqref{eq::agf_eff_p}, the average gate fidelity can be then estimated by the following formula,
\begin{align}\label{f_est}
  \Bar{F} = \prod_{q \in \mathbb{Q}}\frac{1+P_q^{2\lfloor\frac{N}{2}\rfloor}}{2},
\end{align}
where the $\mathbb{Q}$ is the set of qubits representing a QPU. %We can treat this as the terms of the original random circuit, so if one had a circuit with limited connectivity, then the additional gates are needed hence we need to do some modification on $P_q$. It can be done by introducing the allocation matrix $A_{q,q'}$, such that

\bigskip

For a QPU with limited connectivity, additional swapping gates are needed.
For multi-QPU DQC, one needs auxiliary qubits for telegating.
Both processes introduce additional noises to the implementation of each SU(4) gate in each layer.
To account
To assess the average AGF of multi-QPU DQC with limited connectivity one needs to incorporate the noises introduced by swapping gates and the telegating processes into the effective preserving factors $\wp_{q}$.
We achieve this by introducing a noise propagation matrix $\{A_{q,q'}\}_{q,q'\in\mathbb{Q}}$ modeled with the single-qubit depolarizing noises.
Each element $A_{q,q'}$ of the noise propagation matrix describes the average number of single-qubit depolarizing channels propagating from the qubit $q'$ to the qubit $q$, when one implements a random two-qubit SU(4) gate on $q$.
One can then adjust the preserving factor $\bar{P}_q$ on the qubit $q$ for each SU(4) gate according to the noise propagation matrix as follows
\begin{align}\label{modifiy error}
  \bar{P}_{q}(A) = \prod_{q'\in \mathbb{Q}}P_{q'}^{A_{q,q'}},
\end{align}
where $\mathbb{Q}$ is the set of all qubits on the composited QPUs.
Note that, the preserving factor $P_{q'}$ quantifies the quality of a direct two-qubit gate acting on two connected qubits, while the noise propagation matrix $A_{q,q'}$ characterizes the connectivity topology and the entanglement links of a multi-QPU DQC device.
The effective preserving factor $\vec{\wp}(A)$ that adopts the topology of a multi-QPU DQC with limited connectivity described by the noise propagation matrix $A$ is then given by
\begin{equation}
  \wp_{q}(A) = \bar{P}_{q}(A)^{2\lfloor\frac{N}{2}\rfloor}.
\end{equation}

The connectivity topology of a multi-QPU DQC device can be represented by an extended connectivity graph.
For example, in Fig.~\ref{fig_3d1-graph-table} (b) and (c), a straight-line edge between two qubits represents the ability of direct implementation of a two-qubit gate on a single QPU, while a wiggly-line edge between two memory auxiliary qubits represents a quantum channel establishing a pair of entangled qubits.
The complete set of qubits $\mathbb{Q} = \mathbb{Q}_{w}\cup\mathbb{Q}_{m}$ includes the set of working qubits $\mathbb{Q}_{w}=\{q_{1},..., q_{8}\}$ and extended by the set of memory qubits $\mathbb{Q}_{m}=\{m_{A},m_{B}\}$.
Since the benchmarking only evaluates the measurement readouts of the working qubits, the average AGF is solely determined by the effective preserving factors $\wp_{q}$ of the working qubits $q \in \mathbb{Q}_{w}$. One therefore arrives at the following formula for the average AGF in a multi-QPU DQC device,
%\begin{align}
%\Bar{F} &= \prod_{q \in \mathbb{Q}_{w}}\frac{1+\Bar{P}_q^{2\lfloor\frac{N}{2}\rfloor}}{2}.
%\label{f_est_g}
%\end{align}
\begin{align}
\Bar{F}(A) &= \prod_{q \in \mathbb{Q}_{w}}\frac{1+\wp_{q}(A)}{2}.
\label{f_est_g}
\end{align}

\bigskip

\begin{figure*}
  \centering
  \includegraphics[width=\textwidth]{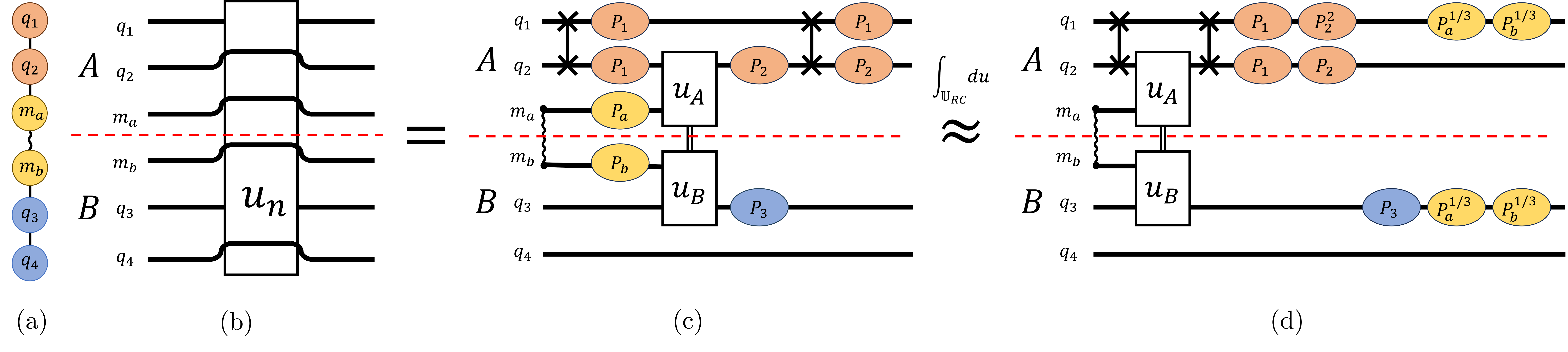}
  \caption{From extended connectivity graph to cost matrices. (a) An extended connectivity graph with the working qubits $\mathbb{Q}_{w} = \{q_{1},q_{2},q_{3},q_{4}\}$ and the memory auxiliary qubits $\mathbb{Q}_{m} = \{m_{a},m_{b}\}$.
  (b) A random two-qubit gate $u_{n}$ acts on the qubit $\{q_{1},q_{2}\}$.
  (c) An implementation of $u_{n}$ with single-qubit depolarizing noises.
  (d) Moving all single-qubit depolarizing noises to the end of the circuit.
  }\label{fig::cost_matrix}
\end{figure*}

To determine the noise propagation matrix for a given extended connectivity graph, we analyze the implementation cost of a random two-qubit gate between two qubits $q$ and $q'$.
The cost that one needs to pay is the single-qubit depolarizing channels introduced by additional swapping gates under limited connectivity of $\mathbb{Q}_{w}$ and telegating across the auxiliary qubits $\mathbb{Q}_{m}$.
For the example shown in Fig. \ref{fig::cost_matrix}, in the $n$-th layer of a QV random-circuit benchmarking, one randomly chooses a two-qubit unitary $u_{n}$ acting on the qubits $\{q_{1},q_{3}\}$, as shown in Fig. \ref{fig::cost_matrix} (b).
Under the extended connectivity of the two-QPU DQC device shown in Fig. \ref{fig::cost_matrix} (a), the implementation of $u_{n}$ introduces depolarizing channels, which are depicted as the circles in Fig. \ref{fig::cost_matrix} (c).
Taking the average over the QV random-circuit sampling, the one can approximately moving all single-qubit depolarizing channels to the end of circuit with the effective preserving factors shown in Fig. \ref{fig::cost_matrix} (d).
For the implementation of $u_{n}$ on $\{q_{1},q_{3}\}$ under the extended connectivity graph, one can a cost matrix $C(q_{1},q_{3})$ as follows,
\begin{equation}
\label{eq::cost_matrix_eg}
  C(q_{1},q_{3})
  =
  \begin{bmatrix}
    1 & 2 & 0 & 0& \frac{1}{3} & \frac{1}{3}\\
    1 & 1 & 0 & 0& 0 & 0 \\
    0 & 0 & 1 & 0&  \frac{1}{3} & \frac{1}{3} \\
    0 & 0 & 0 & 0&  0 & 0
  \end{bmatrix}.
\end{equation}
Each element $C_{n,m}(q_{1},q_{3})$ of the cost matrix counts the number of depolarizing channels originated from the qubit $q_{m}$ and acting on the working qubit $q_{n}$.
Note that a cost matrix $C(q,q')$ contain $N$ rows of working qubits and $N+M$ columns of working and auxiliary memory qubits.
The detailed steps and examples of calculating this cost matrix are provided in Appendix
\ref{app_exp cost}.

Finally, we define the noise propagation matrix $A$ by aggregating the cost matrices of all pairs of working qubits,
\begin{align}
\label{eq::allo_matrix}
A \equiv \frac{1}{2(N-1)}\sum_{q,q'\in \mathbb{Q}_w} C(q,q')
\end{align}
Here, the cost matrices are divided by $2(N-1)$ to ensure that the noise propagation matrix of a fully connected QPU is normalized to $A_{q,q'} = \delta_{q,q'}$, such that Eq.~\eqref{f_est} aligns with Eq.~\eqref{f_est_g} as a special case.

\section{Numerical simulation of benchmarking}\label{sec::num_sim_RB}
In this section, we employ numerical simulation to scrutinize the one-to-one correspondence among HOP, LXE, and AGF in our single-qubit depolarizing approximation outlined in the preceding section. We adopt the average AGF as the figure of merit for the two-QPU DQC benchmarking, and show a performance improvement of DQC under limited connectivity, which indicates the enhancement of scalability of quantum computing through DQC.

\bigskip

We employ the same random circuit structure utilized in quantum volume benchmarking to compute both the HOP and linear cross-entropy.
However, it should be noted that the specific random circuit structure chosen should not undermine our results, given that they all effectively approximate the Haar uniform unitaries.
To reduce the total number of gates and the efficiency of the circuit compiling, we replace the permutation gates in QV random circuits by repositioning the SU(4) gates.
The SU(4) gates are samples through the KAK decomposition, in which any SU(4) gate can be expressed as a sequence of single-qubit operations and three CNOT gates.

\begin{figure}[htb]
    \centering
    \includegraphics[width=0.48\textwidth]{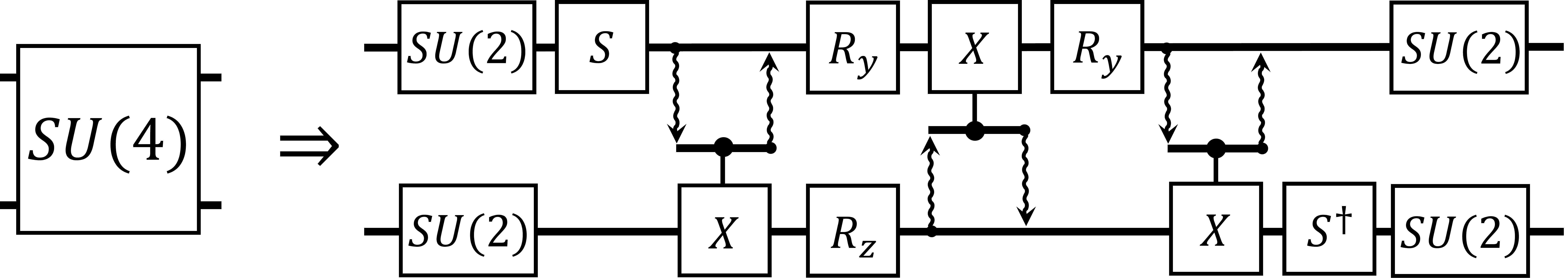}
    \caption{Implementation of a non-local SU(4) gate using the EJPP protocol. The diagram illustrates the application of three independent EJPP protocols to realize the SU(4) gate via DQC. }
    \label{fig_kak}
\end{figure}

For global SU(4) gates acting across the two local QPUs, we apply the EJPP protocol to these CNOT gates. We then can implement all the non-local SU(4) gates through DQC, as depicted in Fig. \ref{fig_kak}.
It is worth noting that this approach necessitates three entangled pairs to implement a single SU(4) gate.
Alternatively, utilizing quantum state teleportation reduces this requirement to two entangled pairs (one for sending the state and one for returning it).
However, adopting quantum state teleportation entails the addition of one more qubit to each local QPU.
The utility of the EJPP telegating protocol minimizes the number of auxiliary memory qubits to one within each local QPU, and hence maximize the number of working qubits.

\begin{figure}[htb]
    \centering
    \includegraphics[width=0.45\textwidth]{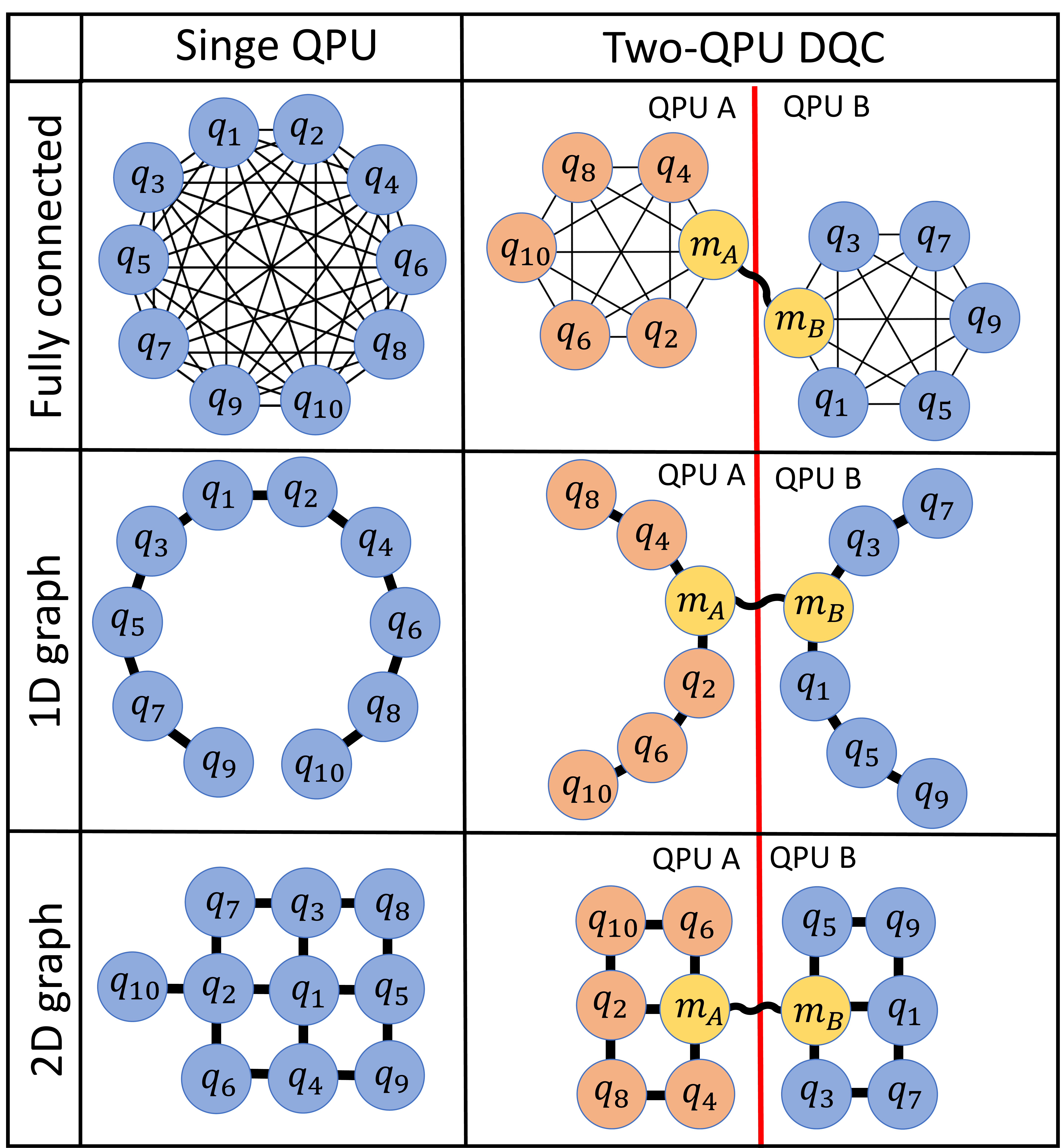}
    \caption{Extended connectivity graphs for simulation.
    The first column is the group of single-QPU devices, where the blue vertices represent the working qubits that are responsible for the computing, and the black edges identify the direct coupling between qubits.
    The second column is the group of two-QPU DQC devices, where the blue and orange vertices represent the working qubits of the local QPUs, and the yellow vertices denote the auxiliary memory qubits.
    Each auxiliary memory qubit can either store the shared entangled state or facilitate the passage of additional swapping gates. }
    \label{fig_g_table}
\end{figure}

In our simulation, we select six types of connectivity, which can be categorized into two groups.
The first group comprises devices utilizing a single QPU, while the second group consists of DQC-composited devices equipped with two local QPUs.
Within each group, the local QPUs exhibit three types of connectivity graphs, namely fully connected graph, 1D (line) graph, and 2D (grid) graph, respectively.
Fig.~\ref{fig_g_table} illustrates the extended connectivity graphs of our simulation devices. Solid lines denote the coupling between physical qubits, while wavy lines represent the sharing of entangled pairs between auxiliary memory qubits.
For the two-QPU DQC devices, each local QPU contains only one auxiliary memory qubit.
The allocation of the auxiliary memory qubit is chosen as the hub qubit, which improves the connectivity to the greatest extent.
\wjy{[Explain, fix topology for local QPUs]}

\begin{figure*}[htbp]
    \centering
    \hfill
    \subfloat[]{\includegraphics[width=0.43 \textwidth]{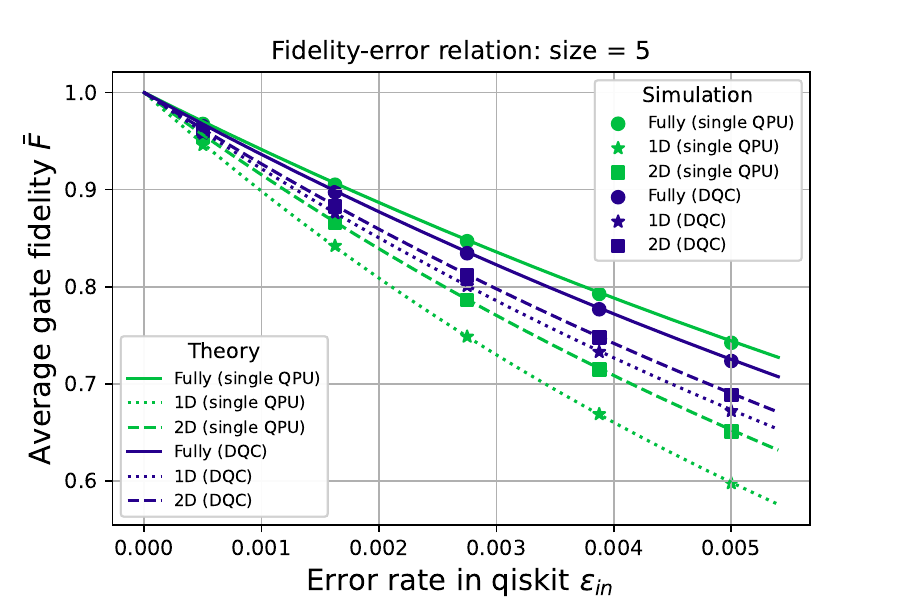}}
    \hfill
    \subfloat[]{\includegraphics[width=0.43 \textwidth]{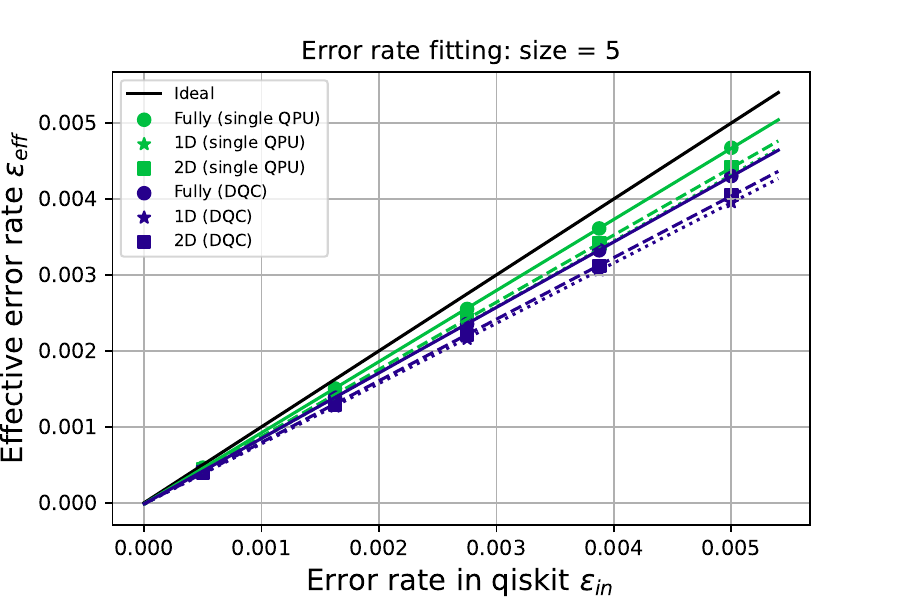}}
    \hfill{  }
    \vspace{-1.5em}
    \\
    \hfill
    \subfloat[]{\includegraphics[width=0.42 \textwidth]{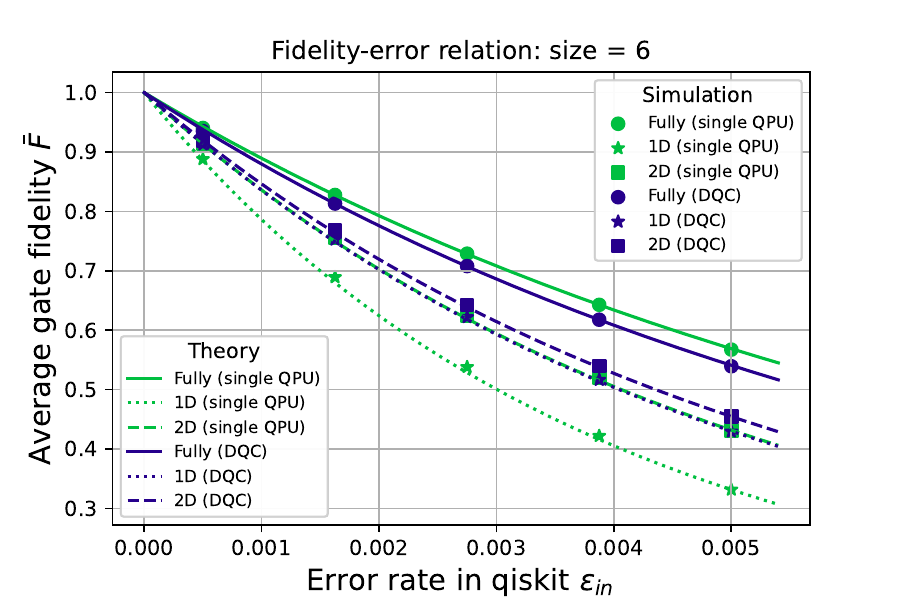}}
    \hfill
    \subfloat[]{\includegraphics[width=0.42 \textwidth]{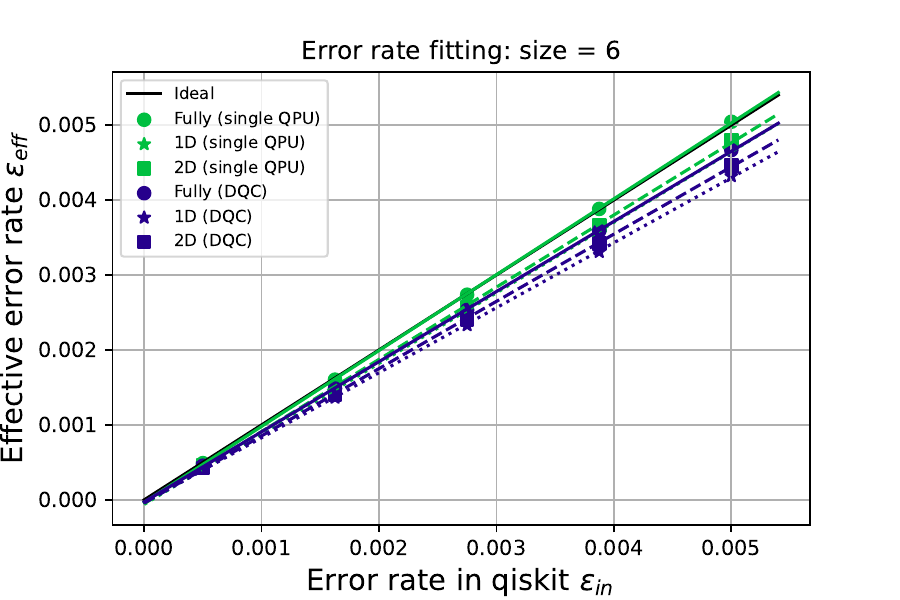}}
    \hfill{  }
    \vspace{-1.5em}
    \\
    \hfill
    \subfloat[]{\includegraphics[width=0.42 \textwidth]{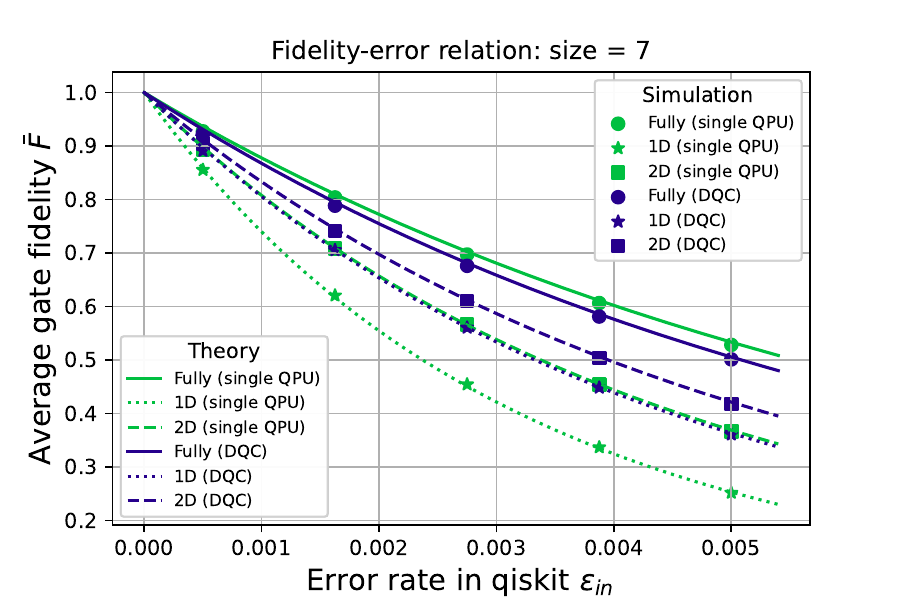}}
    \hfill
    \subfloat[]{\includegraphics[width=0.42 \textwidth]{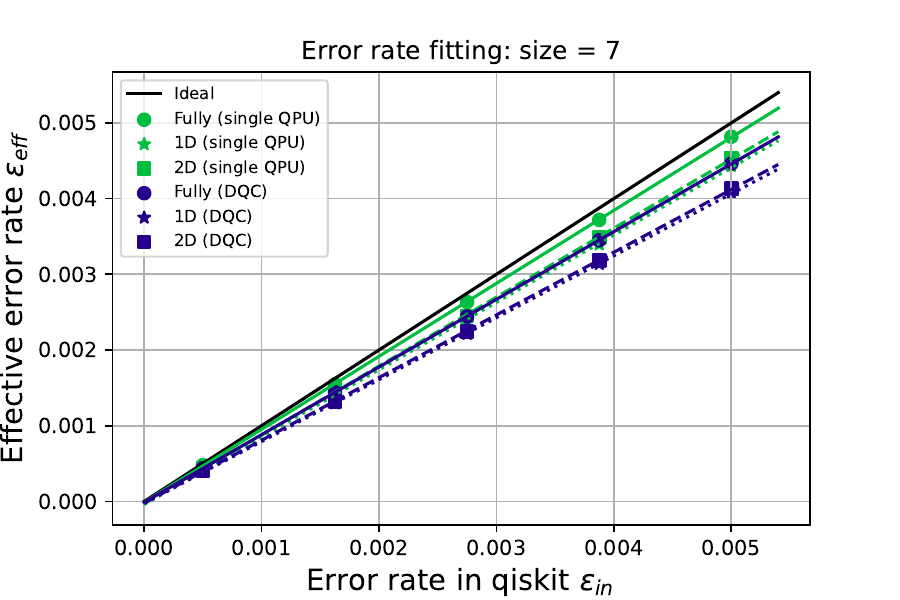}}
    \hfill{  }
    \vspace{-1.5em}
    \\
    \hfill
    \subfloat[]{\includegraphics[width=0.42 \textwidth]{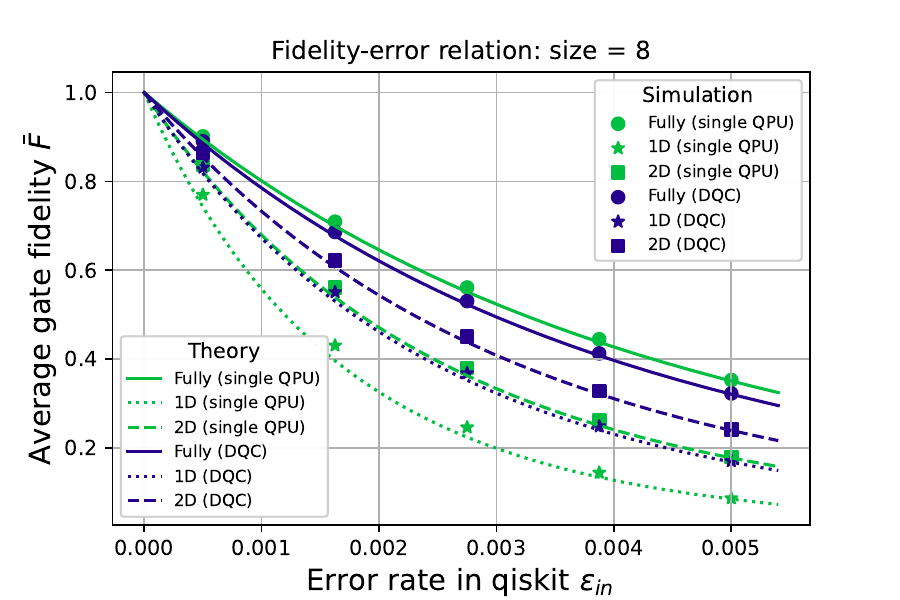}}
    \hfill
    \subfloat[]{\includegraphics[width=0.42 \textwidth]{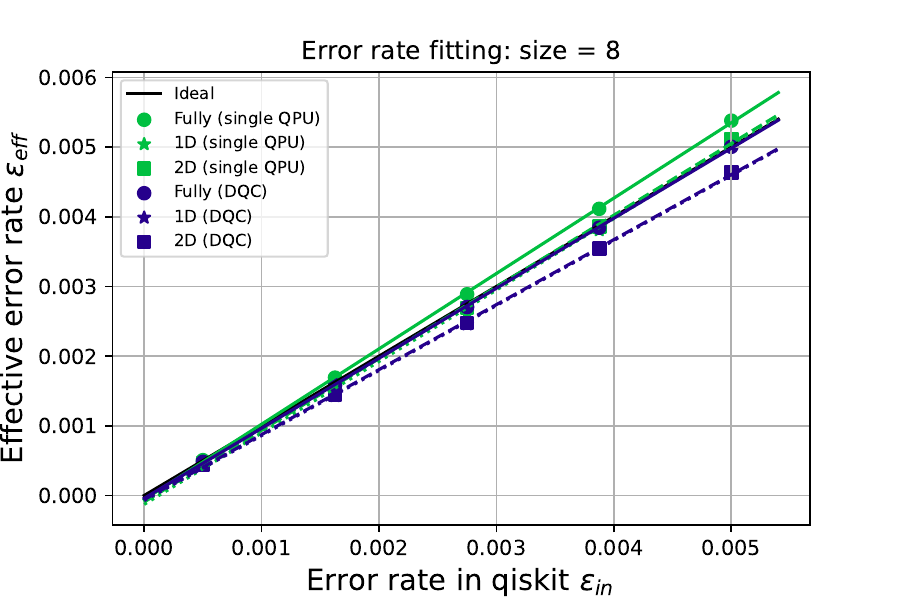}}
    \hfill{  }
    \vspace{-1em}
    \caption{
    (a),(c),(e),(g) Average gate fidelity of $5$-qubit devices with different types of connectivity.
    The data points represent the average gate fidelity obtained from numerical simulation for different error rates.
    The lines show the theoretical prediction of average gate fidelity derived from Eq.~\eqref{f_est_g}.
    (b),(d),(f),(h) Comparison between the error rate specified in the qiskit noise model and the error rate estimated through our average gate fidelity formula in Eq.~\eqref{f_est_g}. The markers represent the data points obtained from our simulation, while the lines represent the linear fitting of our data.
    }
    \label{fig_error-average gate fidelity}
\end{figure*}

We employ qiskit to simulate the QV random benchmarking under the single-qubit depolarizing noise model.
With Eq.~(\ref{xeb-gf}), we employ linear cross-entropy for measuring average gate fidelity.
The simulation is conducted as follows.
In the first run, we set a constant error rate $\epsilon=0.05\%$ ($\epsilon \equiv 1-P$) across all qubits in qiskit, and conduct QV random-circuit sampling on different types of connectivity for $n$-qubit devices, where $n=5,...,8$.
\wjy{For simplicity, we assumed that all entangled pairs were perfect.}
In the next runs, we vary the error rates $\epsilon$ from $0.05\%$ to $0.5\%$ and implement the same QV random-circuit sampling.
We sampled over $1000$ random circuits for each connectivity graph and error rate and executed $10,000$ shots for each sampled circuit.
After the sampling, for each connectivity graph and error rate, we obtain an average AGF $\bar{F}$, which are plotted by different markers in Fig.~\ref{fig_error-average gate fidelity} (a,c,e,g).

To verify the result in Eq.~\eqref{f_est_g}, we calculate the effective preserving factor $P_{\text{eff}}$ of a quantum gate, as well as the effective error rate $\epsilon_{\text{eff}} \equiv 1 -  P_{\text{eff}}$ from each average AGF data according to Eq.~\eqref{f_est_g}.
In Fig.~\ref{fig_error-average gate fidelity} (b,d,f,h), the effective error rates $\epsilon_{\text{eff}}$ are compared with the initial inputs $\epsilon_{\text{in}}$ of the qiskit function.
From the simulation, we observe that the effective error rates are linearly proportional to the input error rates with a ratio $r_{G,n}$ around $0.8\sim 1$ varying with different types of connectivity $G$ and qubit numbers $n$,
\begin{equation}
  \epsilon_{\text{eff}} = r_{G,n} \epsilon_{\text{in}}.
\end{equation}
Ideally, one expects $r_{G,n}=1$, however, a small distinction arises.
This discrepancy stems from the difference between our error model and the compilation in qiskit.
In our error model, we treat every SU(4) gate as a fundamental building block and assume that all the errors are single-qubit depolarizing noises.
On the other hand, in qiskit, an SU(4) gate is decomposed into a sequence of single-qubit gates and CNOT gates, which are subject to single-qubit and two-qubit depolarizing channels, respectively.
\wjy{The strong linearity between $\epsilon_{\text{eff}}$ and $\epsilon_{\text{in}}$ provides a compelling evidence for the validity of Eq.~\eqref{f_est_g}.}

Due to the lack of knowledge about the detailed compilation in qiskit, we should use the ratio $r_{G,n}$ fitted in Fig.~\ref{fig_error-average gate fidelity} (b,d,f,h) to calibrate the effective preserving factor from Eq.~\eqref{modifiy error},
\begin{equation}
  \wp_{q} = \prod_{q'\in\mathbb{Q}} (1-r_{G,n} + r_{G,n} P_{\text{in}})^{A_{q,q'}}.
\end{equation}
Employing this calibrated effective preserving factor, we plot the theoretical prediction of Eq.~\eqref{f_est_g} for different connectivities by lines in Fig.~\ref{fig_error-average gate fidelity} (a,c,e,g).
One can observe that the theoretical predictions in Eq.~\eqref{f_est_g} and the simulate results evaluated through the AGF-HOP correspondence in Eq.~\eqref{hop_gf}, align with each other very well.
It verifies both the AGF-HOP correspondence in Eq.~\eqref{hop_gf} and our error model in Eq.~\eqref{f_est_g}.

\begin{figure}[t]
    \centering
    \includegraphics[width=0.5 \textwidth]{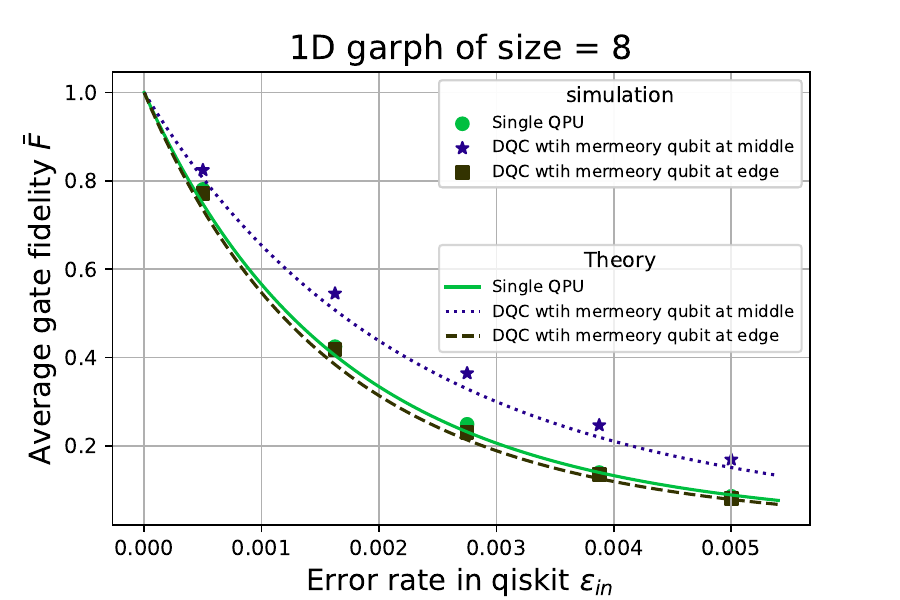}
    \caption{Average AGF of 1D connectivity graph shown in Fig.~\ref{fig_3d1-graph-table} for different error rates.
    The two-QPU DQC device in Fig.~\ref{fig_3d1-graph-table}(b) featuring memory qubits positioned in the middle, which exhibits a significant advantage over the single-QPU device. However, the two-QPU DQC device with memory qubits at the edges does not offer any advantage. Therefore, the strategic selection of the auxiliary memory qubit's positions on each local QPU emerges as a critical condition for scalability enhancement.
    }
    \label{fig_3d1-exp}
\end{figure}

Besides, in Fig.~\ref{fig_error-average gate fidelity} (a,c,e,g), one can also observe that a single fully connected QPU exhibits the highest average gate fidelity among the different connectivity graphs, which is expected as there are no additional gates introduced. Its corresponding two-QPU DQC devices perform slightly worse, as additional gates for telegating are applied.
On the contrary, for 1D and 2D connectivity, the two-QPU DQC devices outperform the single-QPU device for all $n$-qubit devices, which shows the enhancement of scalability in two-QPU DQC under limited connectivity.
The condition for such scalability enhancement depends on the connectivity of QPUs.
To confirm this assertion, we conduct another simulation for $1D$ connectivity, where we relocate the auxiliary memory qubit in each QPU to the end of the line graph, as illustrated in Fig.~\ref{fig_3d1-graph-table} (c).
In this configuration, the two-QPU DQC device resembles a single-QPU device with a line graph topology.
Since the implementation of DQC does not enhance the connectivity, and meanwhile there will be additional error costs introduced by telegating, we expect no scalability enhancement for this configuration, which aligns with the simulation results plotted in Fig.~\ref{fig_3d1-exp}.

\bigskip

\begin{figure}[t]
    \centering
    \hfill
    \subfloat[]{\includegraphics[width=0.5\textwidth]{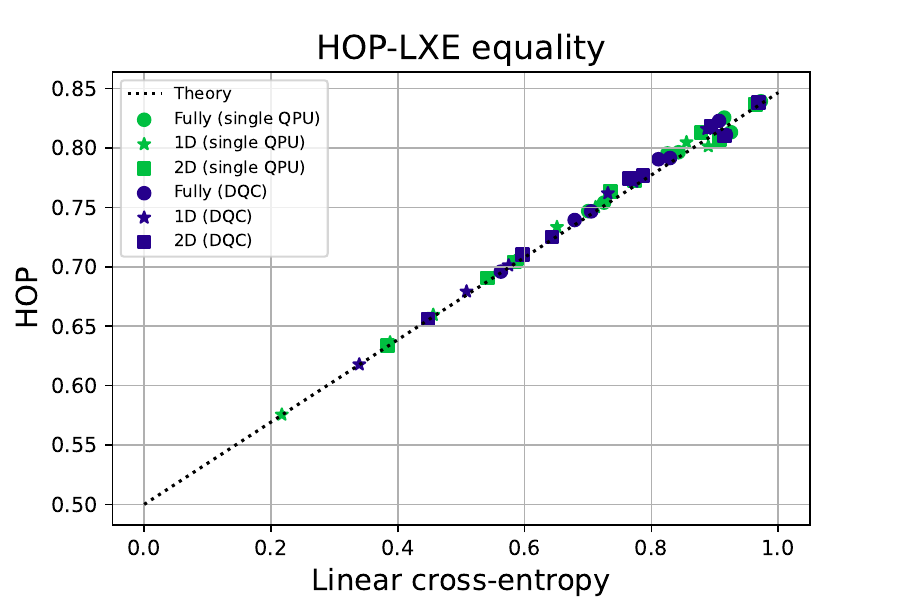}}
    \hfill
    \\
    \hfill
    \subfloat[]{\includegraphics[width=0.5\textwidth]{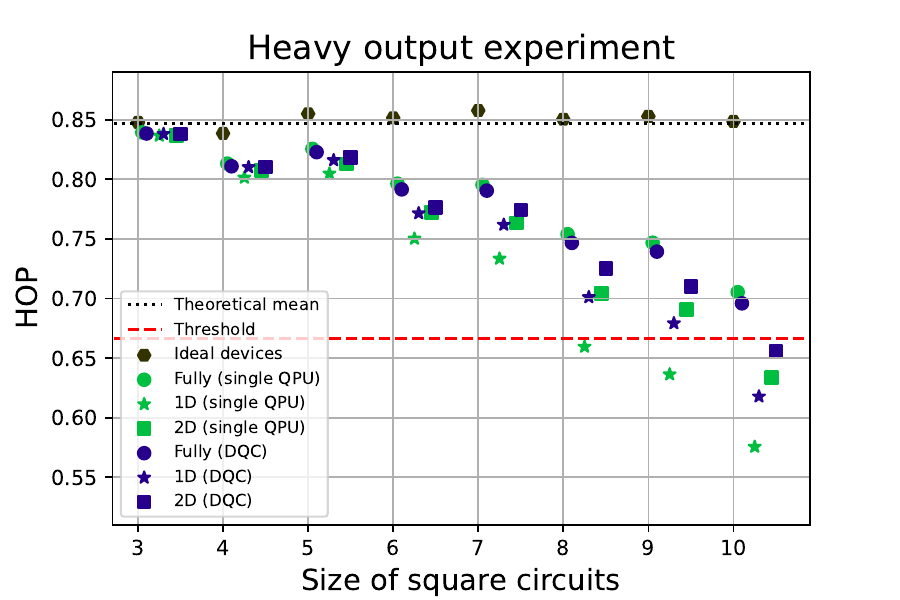}}
    \hfill{ }
    \caption{
    QV random-circuit benchmarking for $n$-qubit devices with different types of connectivity. Here the size of devices ranges from $3$ to $10$.
    (a) HOP-LXE correspondence.
    %The figure illustrates HOP and LXE data points for different sizes and types of devices. The black line corresponds to the theoretically predicted derived from Eq.~(\ref{hop-xeb}). \gb{The data spans from 3 qubits to 10 qubits. Notably, for each type of spot, the closer to the left indicates results from larger-sized circuits.}
    (b) HOP for different sizes and connectivity.
    %The figure depicts the simulation results of the heavy output experiments, \gb{showcasing the behavior of the HOP as the size of devices increases while maintaining a fixed error rate}. The distinct colors \gb{denote} ideal (error-free), single-QPU, and DQC-decomposed devices, \gb{while} different shapes represent different connectivity patterns of local QPUs.
    }
    \label{fig_hop-xeb}
\end{figure}

To further support the AGF-HOP-LXE correspondence and scalability enhancement of DQC, we conduct additional simulations, in which we fix the error rate $\epsilon_{\text{in}}=0.15\%$ for all qubits, and benchmark both single-QPU and two-QPU devices from $3$-qubit to $10$-qubit.
Under this setting, we sample over $2000$ random circuits for all sizes and take $10000$ shots for each random circuit.
For each configuration of connectivity graph $G$ and qubit number $n$, we evaluate the measurement outputs with both average HOP and LXE and plot them in Fig.~\ref{fig_hop-xeb}(a).
One can clearly verify the HOP-LXE correspondence derived in Eq.~\eqref{hop-xeb} with these simulated data as a general property independent of device size and connectivity.
%\gb{Note that, there was a notable exception in the upper right corner of the plot, where the data deviated from the expected trend. This discrepancy was mainly due to including smaller circuit sizes, which our equation wasn't designed to accommodate.}

Employing the obtained data, we evaluate the quantum volume of different devices in Fig.~\ref{fig_hop-xeb}(b), which offers insights into the scalability of device architecture by illustrating how the HOP decreases with increasing device size.
It shows that, for a fully connected QPU, the DQC configuration doesn't offer an advantage over a single-QPU setup.
Note that this statement is only made for the situation when the error rates of multiple small-size QPUs are the same as the error rate of a large-size single QPU.
However, in practice, the error rate of a small-size QPU is usually smaller than the one of a large-size QPU.

Besides the fully connectivity, for the devices with 1D and 2D connectivity graphs, we observe that the two-QPU DQC devices exhibit superior HOP compared to the single-QPU configuration.
Especially for 1D connectivity, we observe the increase of quantum volume from $2^{7}$ to $2^{9}$ through DQC.
These observations suggest that the connectivity of the devices plays a pivotal role in the scalability analysis of multi-QPU DQC under the QV random-circuit benchmarking scheme.

\section{Results and discussion}\label{sec::result}
Our theoretical calculation and numerical simulation of QV random-circuit benchmarking show that the scalability enhancement in DQC strongly depends on the connectivity of QPUs, the choice of the auxiliary memory qubits for establishing a quantum communication channel, \wjy{ and the quality of entanglement distribution.}

\bigskip

First, for fully connected devices shown in Fig.~\ref{fig_error-average gate fidelity} and Fig.~\ref{fig_hop-xeb} (b), DQC does not improve the scaling of quantum computing, if the error rate of a QPU of a larger size is the same as a smaller-size QPU.
The underlying reason is the additional gates in the EJPP protocol that cause additional gate errors and lower the circuit's average gate fidelity.
\begin{equation}
  \bar{F}_{\text{DQC}}
  \underset{\text{fully connected}}{<}
  \bar{F}_{\text{single}}.
\end{equation}
\wjy{Under the same reason, a more general observation is that the multi-QPU DQC configurations that preserve the topology of single QPUs should not have advantage over single-QPU device, if we assume the same error rate of quantum gates on each QPU regardless of their size in qubit numbers.
However, in practice, the gate error rate on a single QPU should increase as the number of qubits increases. Under this consideration, the utility of DQC with small-size QPUs, which have a lower error rate, can also enhance the scalability.
For example, according to benchmarking data is plotted in Fig.~\ref{fig_3d1-exp}, for  the configuration shown in Fig.~\ref{fig::cost_matrix} (c), if the error rate of two $5$-qubit QPUs is $\epsilon = 0.1\%$, one would expect an average AGF around $\bar{F} = 0.55$, which is greater than the average AGF $\bar{F} = 0.43$ of the single-QPU device with $\epsilon = 0.15\%$.
In this regard, even without changes of connectivity topology, one can still expect scalability enhancement.
}

\bigskip

Second, for devices with limited connectivity, DQC can change the connectivity and reduce the number of swapping gates utilized for the implementation of two-qubit gates on two non-neighboring qubits.
To improve connectivity in the best way, the auxiliary memory qubits for sharing an entangled Bell pair should be chosen as the qubit that has the highest degree of neighboring qubits.
Such an enhancement in connectivity can compensate for and even overcome the additional errors introduced by telegating so that one can enhance quantum computing performance by scaling up the system through multi-QPU DQC, as it is shown in Fig.~\ref{fig_error-average gate fidelity} and Fig.~\ref{fig_hop-xeb} (b).
\begin{equation}
  \bar{F}_{\text{DQC}}
  \underset{\text{1D or 2D}}{>}
  \bar{F}_{\text{single}}.
\end{equation}

For large-size QPUs with more complicated connectivity graphs, it is not efficient to try all possible choices of auxiliary memory qubits to find the best DQC configuration for scalability enhancement employing Eq.~\eqref{f_est_g}.
To provide an efficient analytical tool, we introduce further approximation of average AGF for large-size and high-fidelity QPUs,
%\begin{align}
%\Bar{P}_q &= \prod_{q'\in \mathbb{Q}} (1-\epsilon_{q'})^{A_{q,q'}}\approx 1-\sum_{q'\in \mathbb{Q}}A_{q,q'}\epsilon_{q'}\\
%\braket{\epsilon}&=\frac{1}{N}\sum_{q\in\mathbf{Q_w}}(1-\Bar{P}_q)=\frac{1}{N}\sum_{q,q'\in\mathbf{Q_w},\mathbb{Q}}A_{q,q'}\epsilon_{q'}\notag\\
%\Rightarrow\Bar{F} &\approx \prod_{q\in \mathbf{Q_w}}1-\frac{N}{2}\braket{\epsilon}\approx \exp(-\frac{N^2}{2}\braket{\epsilon}) \nonumber\\
%&= \exp(-\frac{N}{2}\sum_{q,q'\in \mathbf{Q_w},\mathbb{Q}}A_{q,q'}\epsilon_{q'})
%\end{align}
\begin{align}\label{final-eq}
  \Bar{F}_\mathbb{Q}(\epsilon)&\approx\exp(-\frac{N\mathcal{A}_\mathbb{Q}}{2}\epsilon),
\end{align}
where the error rate for every qubit is a constant $\epsilon$, and the \emph{characteristic cost} $\mathcal{A}_{\mathbb{Q}}$ of a connectivity configuration $\mathbb{Q}$ is the sum of all elements of the noise propagation matrix $A_{q,q'}$
\begin{equation}
  \mathcal{A}_\mathbb{Q}\equiv \sum_{q\in \mathbb{Q}_w, q'\in\mathbb{Q}} A_{q,q'}.
\end{equation}
Here $\mathbb{Q}_w$ is the set of working qubits and $\mathbb{Q}$ is the set of all qubits including auxiliary qubits for telegating (for detailed derivation see Appendix \ref{sec::apdx_approx_AGF}).
Eq.~(\ref{final-eq}) can be employed as a heuristic method for evaluating connectivity configurations of DQC devices.
Fig.~\ref{fig_final-eq} is the comparison between Eq.~(\ref{f_est_g}) and Eq.~(\ref{final-eq}) for different extended connectivity. and it shows that the approximation works well for $\Bar{F}\gtrapprox 0.6$, which is around the threshold of QV benchmarking. This implies that this approximation is sufficient for evaluating quantum computing performance in QV random-circuit benchmarking.
Moreover, we can also use the approximated formula in Eq.~\eqref{final-eq} to decide the position of the auxiliary memory qubits on local QPU by minimizing the \emph{characteristic cost} $\mathcal{A}_\mathbb{Q}$ over all possible allocation.

\bigskip

\begin{figure}
  \centering
  \includegraphics[width=0.48\textwidth]{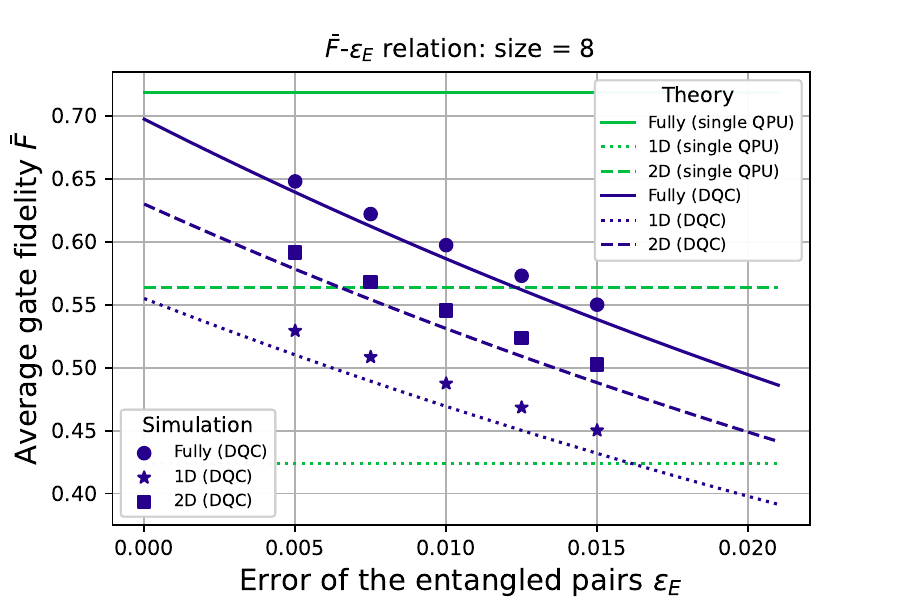}
  \caption{Robustness of scalability enhancement against the imperfection in entanglement.}\label{fig::fig_vs_ent_error}
\end{figure}

\wjy{
Third, there is another implicit requirement for scalability enhancement, that is high-quality entangled pairs shared between local auxiliary memory qubits.
In previous analysis and simulation, we assume perfect entangled pairs are distributed among QPUs and only the noises of local gates involved in the EJPP processes contribute to errors in telegating.
However, the fidelity of entangled pairs strongly affects the capability of telegating to improve the connectivity.
If the entanglement is too noisy, telegating will not be capable of enhancing the scalability.
With a small extension (see Appendix \ref{sec::ent_error_ext}), the theory of fidelity estimation in Eq.~(\ref{f_est_g}) can be employed to analyze the robustness of scalability enhancement in a DQC configuration against the noise $\epsilon_{E}$ in entanglement, where the noise is described as the depolarizing channel with an error rate $\epsilon_{E}$ and a preserving factor $p_{E} = 1-\epsilon_{E}$,
\begin{equation}
  D_{p_{E}}(\Phi) =
  (1-\epsilon_{E}) \projector{\Phi} +  \epsilon_{E}\frac{1}{4}\widehat{\id},
\end{equation}
and $\ket{\Phi}$ is one of the Bell states.
In Fig.~\ref{fig::fig_vs_ent_error}, we simulate the effect of noise in entanglement on the average gate fidelity and compare them with the theoretical prediction.
Our theory predicts that the DQC scalability enhancement for $1D$ and $2D$ connectivity is robust against the noise $\epsilon_{E}$ up to
$0.65\%$ and $1.62\%$, respectively, which are lower bounds on the simulated results.
We can therefore determine the essential requirement on entangled pairs for DQC scalability enhancement employing Eq.~(\ref{f_est_g}).
}

%\begin{figure*}[htb]
%    \centering
%    \subfloat[DQC with size = 4]{\includegraphics[width=0.45\textwidth]{approx 6_0}}
%    \subfloat[DQC with size = 6]{\includegraphics[width=0.45\textwidth]{approx 6_1}}
%    \\
%    \subfloat[single QPU with size = 8]{\includegraphics[width=0.45\textwidth]{approx 8_0}}
%    \subfloat[DQC with size = 8]{\includegraphics[width=0.45\textwidth]{approx 8_1}}
%    \\
%    \subfloat[single QPU with size = 10]{\includegraphics[width=0.45\textwidth]{approx 10_0}}
%    \subfloat[DQC with size = 10]{\includegraphics[width=0.45\textwidth]{approx 10_1}}
%    \caption{The lighter curve stands for the original function, and the darker curve stands for the function under approximation. For size 6, although the size is not large, still we had a good match and the high-fidelity region ($\Bar{F}\gtrapprox 0.6$). For the larger size, the approximation gets closer to the original formula.}
%    \label{fig_final-eq}
%\end{figure*}
%
%

\section{Conclusion}\label{sec::conclusion}
\wjy{In this work, we employ average gate fidelity, heavy output probability, and linear cross-entropy as the figures of merits to benchmark the performance of DQC through random circuit sampling, and compare them with single-QPU quantum computing.
Under an approximation using a single-qubit depolarizing error model, we show the one-to-one correspondence among these three figures of merits in Eq.~\eqref{hop_gf}, \eqref{xeb-gf}, and \eqref{hop-xeb}.
This relation allows us to obtain a unified evaluation of the performance of quantum computing devices in the random-circuit benchmarking.}

\wjy{
The theoretical evaluation according to our theory in Eq. \eqref{f_est_g} agrees with the simulation results, which show the scalability enhancement in DQC for QPUs with limited connectivity, such as 1D and 2D graphs shown in Fig.~\ref{fig_error-average gate fidelity} and \ref{fig_hop-xeb}.}
The results suggest that the condition of scaling up quantum computing using DQC is to choose a good qubit that is connected to a large number of neighbors as the auxiliary memory qubit on each local QPU for the quantum communication channel. And for given local QPUs, we can find the best position for the auxiliary memory qubit, with the help of the noise propagation matrix $A$ introduced in Eq.~\eqref{eq::allo_matrix}, and a heuristics method of finding the best position is through the minimization of the characteristic value $\mathcal{A}_{\mathbb{Q}}$.

\bigskip

In the future, one may generalize the theory in three directions. The first is to consider multipartite DQC-composited devices with more than two local QPUs.
\wjy{The second is to increase the number of auxiliary memory qubits for the quantum communication channel on each local QPU. }
\wjy{For example, if there are two auxiliary memory qubits on each local QPU, the DQC-composed device can support the advanced embedding-enhanced distributing processes proposed in \cite{WuEtAlMurao2023-EntEffDQC, MartinezEtAlDuncan2023-DQCinQNet}. The embedding-enhanced distributing techniques can enhance the efficiency of quantum communication channels and hence improve the average gate fidelity of the telegating implementation.}
\wjy{The last is to incorporate qubit allocation in the random circuit benchmarking, which will reduce the number of swapping gates and entangled pairs, and hence improve the quantum computing performance.
These three directions are highly related to each other and all play important roles in the development of DQC.}

\begin{acknowledgments}
\wjy{This work is supported by NSTC under the Grant No. 111-2923-M-032-002-MY5, 112-2112-M-032-008-MY3, 112-2811-M-032-005-MY2, 112-2119-M-008-007, and 113-2119-M-008-010.}
\end{acknowledgments}

\appendix
\section{The commutative hypothesis}\label{app_comm_hy}
%In section \ref{three_two}, we had a hypothesis that the SU(4) gate is commutable with the single-depolarizing channel. This hypothesis is based on the fact that the random SU(4) gate under the Haar measure is commutable with the single-qubit depolarizing channel. i.e.:
In Section \ref{three_two}, we \gb{proposed} a hypothesis \gb{asserting the commutativity of} the SU(4) gate with the single-qubit depolarizing channel. This hypothesis \gb{stems from the observation} that, under the Haar measure, a random SU(4) gate \gb{tends to} commute with the single-qubit depolarizing channel, \gb{expressed as:}
\begin{align}
    \int_{U\in SU(4)} dU\;\;[\hat{U}, \hat{\mathbb{I}}\otimes \widetilde{D}_P^{(1)}] = 0\;\;\forall P\in[0,1]
\end{align}
%To prove this, we need the important lemma from the \cite{10.1063/5.0038838}, which states that for all linear operators in $\mathcal{H}^d$, the average unitary transformation under the Haar measure is given by:
\gb{To substantiate this claim,} we \gb{draw upon an essential} lemma from reference \cite{KukulskiEtAlZyczkowski2021-RndmQChnnl}, which states that the average unitary transformation under the Haar measure for all linear operators in $\mathcal{H}^d$ is:
\begin{align}
\int_{U\in SU(d)} dU\;UXU^\dag = \frac{tr(X)}{d}\mathbb{I}_d
\end{align}
% Then by vectorization, we rewrite this equation as:
By vectorization, we \gb{reformulate this lemma} as:
\begin{align}
\int_{U\in SU(d)} dU\;\hat{U}|X\rangle\rangle &= \frac{tr(X)}{d}|\mathbb{I}_d\rangle\rangle
\nonumber\\
& = \frac{1}{d}|\mathbb{I}_d\rangle\rangle \langle\langle \mathbb{I}_d|X\rangle\rangle\;\;\forall|X\rangle\rangle
\end{align}
In the vectorized space, the two-qubit Haar random sampling is a projection onto the completely depolarizing channel
\begin{align}
\int_{U\in SU(d)} dU\;\hat{U} = \frac{1}{d}|\mathbb{I}_d\rangle\rangle \langle\langle \mathbb{I}_d|
\end{align}
%Hence for the case of the two-qubit, we have:
Consequently, for the two-qubit \gb{scenario}, we have:
\begin{align}
\int_{U\in SU(4)} dU\;\hat{U} & = \frac{1}{4}|\mathbb{I}_4\rangle\rangle \langle\langle \mathbb{I}_4|
\nonumber\\
& %= \widetilde{D}_0^{(2)}
= \widetilde{D}_0^{(1)}\otimes\widetilde{D}_0^{(1)}.
\end{align}
%and because the depolarizing channel had a property that
\gb{Utilizing the commutivity of the} completely depolarizing channels
$\widetilde{D}^{(N)}_{P}.\widetilde{D}^{(N)}_{P'} =\widetilde{D}^{(N)}_{PP'} =\widetilde{D}^{(N)}_{P'}.\widetilde{D}^{(N)}_{P} $, we arrive at
\begin{align}
%[\widetilde{D}_0^{(1)}\otimes\widetilde{D}_0^{(1)},\hat{\mathbb{I}}_2\otimes\widetilde{D}^{(1)}_P] = 0\;\forall P\in[0,1]
%\\
%\Leftrightarrow
\int_{U\in SU(4)} dU\;\;[\hat{U}, \hat{\mathbb{I}}\otimes \widetilde{D}_P^{(1)}] = 0, \;\forall P\in[0,1].
\end{align}
%which completes the proof.
\gb{This} completes the proof.

\section{The effective noise matrix for HOP in Eq. \eqref{hop_markov}}
\label{sec::proof_effective_markov_in_HOP}
Here we derive Eq. \eqref{hop_markov} for calculating HOP under noises.
The average HOP under the QV random-sampling is
\begin{equation}
  \bar{H}(\vec{\wp})
  =
  \bar{\boldvec{h}}^{T}.\int_{U\in\mathbb{U}_{RC}}dU \; \Pi_{U}.\mathcal{D}(\vec{\wp}).\boldvec{p}(U).
\end{equation}
Since the assemble of random unitaries sampled by QV circuits can be decomposed into the union of disjoint subsets $\mathbb{U}(\boldvec{q})$,
\begin{equation}
  \mathbb{U}_{RC} = \bigcup_{\boldvec{q}}\mathbb{U}(\boldvec{q}),
\end{equation}
where $\mathbb{U}(\boldvec{q})$ is the set of unitaries that generate the probability distribution $\boldvec{q}$ up to a permutation $\Pi_{U}$ on ${\{0,1,...,2^{N}-1\}}$,
\begin{equation}
  \mathbb{U}_{\boldvec{q}}
  \equiv
  \{U\in\mathbb{U}_{RC}: \exists \Pi_{U} \text{ such that } \Pi_{U}.\boldvec{p}(U) = \boldvec{q}\}.
\end{equation}
The integral over $\mathbb{U}_{RC}$ is then
\begin{equation}
  \int_{U\in\mathbb{U}_{RC}}dU
  =
  \int_{\boldvec{q}}\rho_{\boldvec{q}}d\boldvec{q}\int_{U\in\mathbb{U}_{\boldvec{q}}}dU,
\end{equation}
where $\rho_{\boldvec{q}}$ is the probability density function for sampling a distribution $\boldvec{q}$
\begin{equation}
  \rho_{\boldvec{q}}
  \equiv
  \frac{
    \int_{U\in\mathbb{U}_{\boldvec{q}}}dU
  }{
    \int_{U\in\mathbb{U}_{RC}}dU
  }.
\end{equation}
The average HOP is then
\begin{align}
  \bar{H}(\vec{\wp})
  & =
  \bar{\boldvec{h}}^{T}.\int_{\boldvec{q}}\rho_{\boldvec{q}}d\boldvec{q}\int_{U\in\mathbb{U}(\boldvec{q})}dU \; \Pi_{U}.\mathcal{D}(\vec{\wp}).\Pi_{U}^{T}.\Pi_{U}.\boldvec{p}(U)
  \nonumber\\
  & =
  \bar{\boldvec{h}}^{T}.\int_{\boldvec{q}}\rho_{\boldvec{q}}d\boldvec{q}\int_{U\in\mathbb{U}(\boldvec{q})}dU \; \Pi_{U}.\mathcal{D}(\vec{\wp}).\Pi_{U}^{T}.\boldvec{q}
  \nonumber\\
  & =
  \bar{\boldvec{h}}^{T}.
  \left(\int_{U\in\mathbb{U}(\boldvec{q})}dU \; \Pi_{U}.\mathcal{D}(\vec{\wp}).\Pi_{U}^{T}\right).
  \left(\int_{\boldvec{q}}\rho_{\boldvec{q}}\boldvec{q}\,d\boldvec{q}\right)
\end{align}
The two integrals are determined as follows.
\begin{align}
  \int_{U\in\mathbb{U}(\boldvec{q})}dU \; \Pi_{U}.\mathcal{D}(\vec{\wp}).\Pi_{U}^{T}
  & =
  \frac{1}{2^{N}!}\sum_{\Pi \in \mathbf{S}_{2^N}}\Pi^T \mathcal{D}(\vec{\wp})\Pi.
  \nonumber\\
  & \equiv
  \bar{\mathcal{D}}(\vec{\wp}),
\end{align}
and
\begin{equation}
  \int_{\boldvec{q}}\rho_{\boldvec{q}}\boldvec{q}\,d\boldvec{q}
  =
  \int_{U\in \mathbb{U}_{RC}} dU\; \Pi_U. \boldvec{p}(U).
\end{equation}
As a result, the average HOP is
\begin{equation}
  \bar{H}(\vec{\wp}) = \bar{\boldvec{h}}^{T}.\bar{\mathcal{D}}(\vec{\wp}).\bar{\boldvec{p}}
\end{equation}

After averaging over permutations, the diagonal element of $\Bar{\mathcal{D}}(\vec{\wp})$ is the average of all diagonal elements of $\mathcal{D}(\vec{\wp})$, while the off-diagonal elements of $\Bar{\mathcal{D}}(\vec{\wp})$ is the average of all off-diagonal elements of $\mathcal{D}(\vec{\wp})$.
The diagonal elements in $\Bar{D}(\vec{P})$ is therefore equal to $\Bar{F}$.
Since $\Bar{D}(\vec{P})$ is a symmetric Markov matrix,
%ensuring the sum of elements equals 1 for all rows and columns,
we deduce that all off-diagonal elements are equal to $\frac{1-\Bar{F}}{2^N-1}$.
\begin{align}
  \Bar{\mathcal{D}}(\vec{\wp}) &=
  \begin{bmatrix}
    \Bar{F} & \frac{1-\Bar{F}}{2^N-1} & \dots  & \frac{1-\Bar{F}}{2^N-1} \\
    \frac{1-\Bar{F}}{2^N-1} & \Bar{F}  & \dots  & \frac{1-\Bar{F}}{2^N-1} \\
    \vdots & \vdots  & \ddots & \vdots \\
    \frac{1-\Bar{F}}{2^N-1} & \frac{1-\Bar{F}}{2^N-1} & \dots  & \Bar{F}
  \end{bmatrix}.
\end{align}

\section{Cost matrices for noise propagation matrices}\label{app_exp cost}

For the swapping gates, we introduce two single-qubit depolarizing channels after each swapping gate.
As is customary, we aim to position depolarizing channels at the end of the gate sequence.
Although swapping gates do not commute with single-qubit depolarizing channels, a swapping gate after a single-qubit depolarizing channel moves the noise to another qubit, which allows us to directly position the depolarizing channels at the end of the circuit.
Note that one may reduce the number of applied swapping gates if one does not swap the qubits back to their original place. However, in our analysis, we adopt the approach that one immediately swaps the qubits back to their original position after the implementation of the two-qubit gate as shown in Fig. \ref{con_swap}.
This will keep the qubit allocation unchanged and let the additional swap gates form a closed swapping path.

For a multi-QPU DQC device, we incorporate a non-local telegate employing the EJPP protocol.
In that case, we need to transfer the effects of noises from auxiliary qubits to the working qubits.
To this end, We introduce single-qubit depolarizing channels on both sides of the working qubits, characterized by a preserving factor of $(P_{a}P_{a'})^{\frac{1}{3}}$, where $P_{a}$ and $P_{a'}$ represent the preserving factors of the auxiliary qubits $a$ and $a'$, respectively.
The proof of this noise-shifting phenomenon is provided in Appendix~\ref{app_error shift}.

%For the devices with M auxiliary qubits and N working qubits, the cost matrix $C(q,q')$ for a gate can be given through the following steps:
For devices \gb{comprising} $M$ auxiliary qubits and $N$ working qubits, the cost matrix $C(q,q')$ of a two-qubit gate $U(q,q')$ acting on two qubits $(q,q')$ describes the single-qubit depolarizing channels on each qubit and its origin qubits. It can be \gb{determined} through the following steps:
\begin{enumerate}
    \item \gb{Identify} the swapping path \gb{as per} the device compiler.%Find the swapping path according to the compiler of the devices.
    \item For the swapping path that involves the EJPP protocol, \gb{eliminate the depolarizing channel on the auxiliary qubits and introduce depolarizing channels with an} error parameter %we can remove the depolarizing channel at the auxiliary qubits, by adding the depolarizing channel with the error parameter
    $(P_aP_{a'})^{\frac{1}{3}}$ \gb{on} both sides of the working qubit, \gb{where $P_a$ and $P_{a'}$ represent the error parameters} of the auxiliary qubits.
    \item Put all the depolarizing channels at the end of the swapping path.
    \item The \gb{resulting} cost matrix is an $N+M$ by $N+M$ matrix, where the component $C_{n,m}(q,q')$ \gb{denotes} the number of the depolarizing channel with an preserving factor $P_{m}$ acting on the qubit $n$.
        Fractions are permissible in the case of non-local gates that require telegating.
\end{enumerate}
%Here we give two examples of calculating the cost matrix for a given swapping path. Figure~\ref{fig_swapping-path} shows the treatment of calculating the cost matrix for the local gate, and the corresponding cost matrix is given by Eq.(\ref{ex-cost}).

\begin{figure}[htb]
    \centering
    \includegraphics[width = 0.5\textwidth]{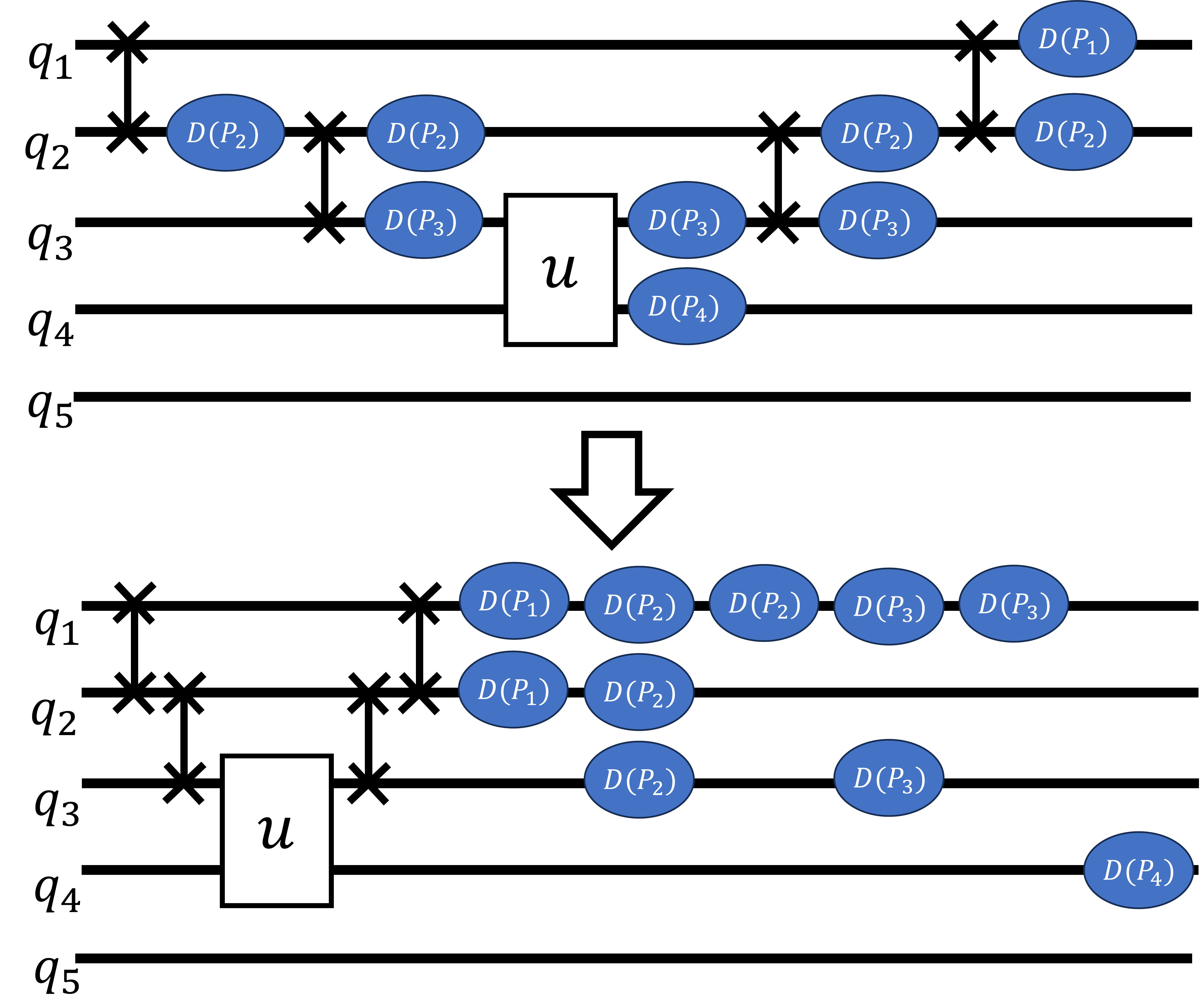}
    \caption{%The upper figure demonstrates a swapping path from qubit 1 to qubit 4, given by the compiler. The lower figure demonstrates how we put all the depolarizing channels to the end of the swap gate, and by counting the numbers of depolarizing channels on each qubit with the label of different error perimeter, we find the cost matrix $C(1,4)$, which is given by the Eq.(\ref{ex-cost})
    Figure \gb{in the upper panel illustrates} a swapping path from qubit 1 to qubit 4, \gb{as provided} by the compiler, and figure \gb{in the lower panel depicts the consolidation of all depolarizing channels} at the end of the swapping path. By counting the depolarizing channels on each qubit \gb{labeled with} different error parameters, we \gb{determine} the cost matrix $C(1,4)$,  as described by Eq.~(\ref{ex-cost}).}
    \label{fig_swapping-path}
\end{figure}

\begin{figure}[htb]
  \centering
  \includegraphics[width = 0.5\textwidth]{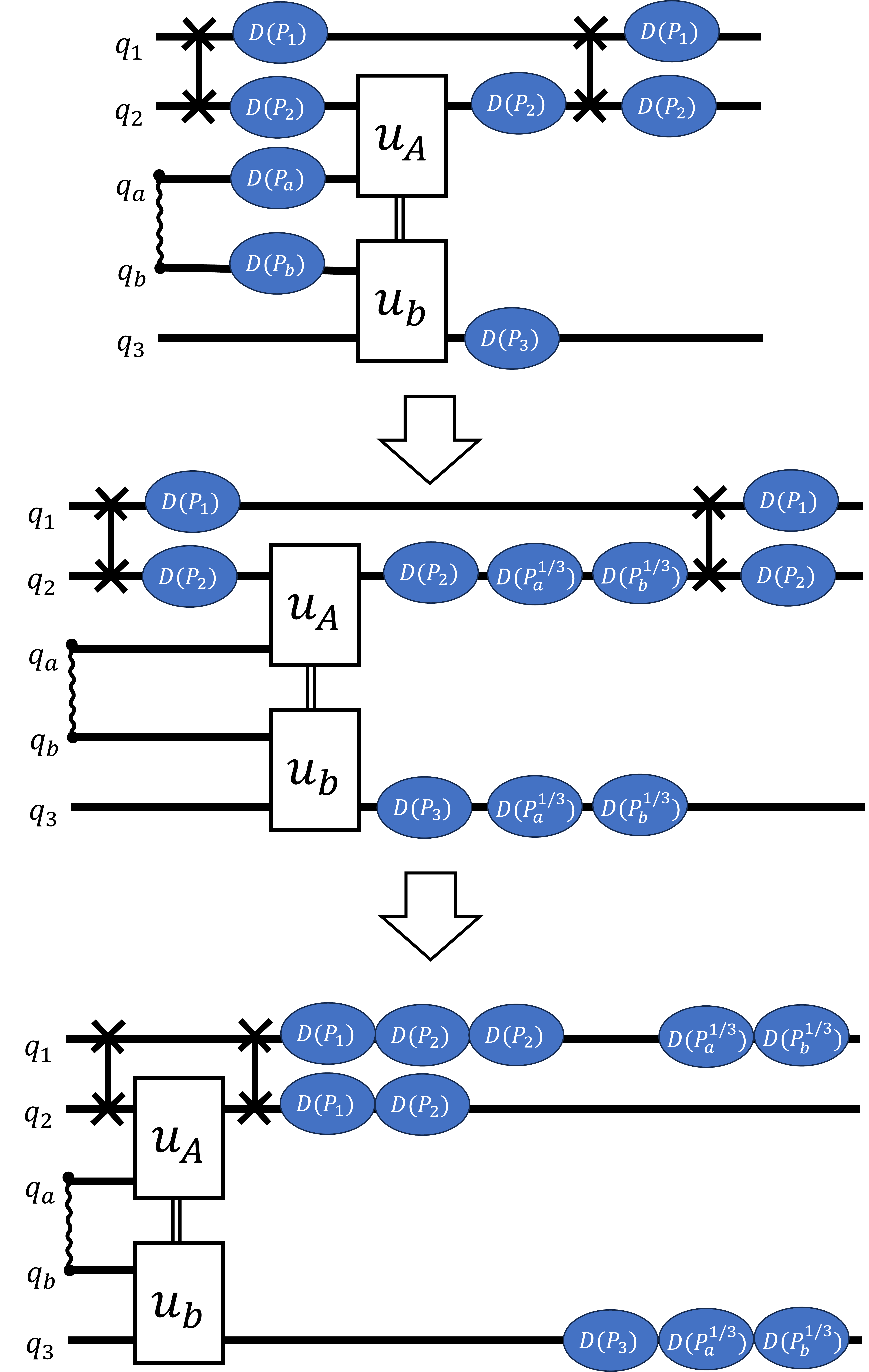}
  \caption{%The upper figure demonstrates a swapping path from qubit 1 to qubit 3 which involves the EJPP protocol. The middle figure demonstrates the error shifting of the
  The figure in the top panel \gb{illustrates} a swapping path from qubit 1 to qubit 3 involving EJPP protocol, the figure in the middle panel demonstrates the error shifting process \gb{from auxiliary qubits to working qubits, and the figure in the bottom panel depicts the consolidation of all depolarizing channels at the end of the swapping path.}
  Through this process, we \gb{derive} %find
  the cost matrix $C(1,3)$, \gb{as described} by %which is given by the
  Eq.~(\ref{ex-cost-dqc})}
  \label{fig_swapping-path-dqc}
\end{figure}

\gb{To illustrate, we provide} two examples of \gb{computing} the cost matrix for a given swapping path. Figure~\ref{fig_swapping-path} \gb{outlines the process} for calculating the cost matrix for a local gate on a QPU, \gb{with the} corresponding cost matrix \gb{described} by Eq.~\eqref{ex-cost},
\begin{align}\label{ex-cost}
C(1,4) =
\begin{bmatrix}
1 & 2 & 2 & 0 & 0\\
1 & 1 & 0 & 0 & 0\\
0 & 1 & 1 & 0 & 0\\
0 & 0 & 0 & 1 & 0\\
0 & 0 & 0 & 0 & 0
\end{bmatrix}.
\end{align}
Fig.~\ref{fig_swapping-path-dqc} \gb{outlines the process for} %shows the treatment of
calculating the cost matrix for a non-local gate across two QPUs, with the corresponding cost matrix \gb{described} %is given
by Eq.~\eqref{ex-cost-dqc}.
\begin{align}
\label{ex-cost-dqc}
C(1,3)=\begin{bmatrix}
1 & 2 & 0 & \frac{1}{3} & \frac{1}{3}\\
1 & 1 & 0 & 0 & 0 \\
0 & 0 & 1 & \frac{1}{3} & \frac{1}{3}
\end{bmatrix}.
\end{align}
Note that the cost matrix $C(q,q')$ is zero when $q = q'$, which corresponds to no two-qubit gate.

%And of course, the cost matrix is zero if $q = q'$. We defined the allocation matrix A, by the average of the cost matrix of all possible gates.
The noise propagation matrix $A$ is then defined in Eq. \eqref{eq::allo_matrix} as the average of the cost matrix of all possible gates.
%\begin{align}
%A \equiv \frac{1}{2(N-1)}\sum_{q,q'\in \mathbb{Q}_w} C(q,q')
%\end{align}
%The connectivity of the devices and the allocation algorithm of the compiler are all embedded in the allocation matrix $A_{q,q'}$.
The connectivity of the devices and the allocation algorithm of the compiler are \gb{encapsulated within} the noise propagation matrix $A_{q,q'}$.

\section{Error propagation in the EJPP telegating}\label{app_error shift}
%For the depolarizing channel acting on the auxiliary qubits, one can first shift it to the beginning of the circuits, where the entangled state becomes:
\gb{When considering} a depolarizing channel acting on the auxiliary qubits, one can first shift it to the beginning of the circuit, \gb{resulting in} a noisy entangled state,
\begin{align}
\label{w_state}
  \rho_{\Phi^+}(P_e) = P_e|\Phi^+\rangle\langle\Phi^+|+\frac{1-P_e}{4}\mathbb{I}_4,
\end{align}
where the $P_e = P_a P_b$. \gb{Consequently,} the channel of the EJPP protocol circuit \gb{can be} treated as a convex combination with a perfect part %hence we can treat the channel of the EJPP protocol circuit as the convex combination with the perfect part
($|\Phi^+\rangle\langle\Phi^+|$) and a \gb{completely} mixed part % and the completed mixed part
($\frac{1}{4}\mathbb{I}_4$). %The perfect part provides the desired CONT gate and the channel of the completed mixed part is given by:
The perfect part \gb{yields} the desired CONT gate, \gb{while} the channel of the \gb{completely} mixed part is given by
\begin{align}
\widetilde{CX} = (\frac{1}{2}\hat{\sigma}_0 + \frac{1}{2}\hat{\sigma}_3)\otimes(\frac{1}{2}\hat{\sigma}_0 + \frac{1}{2}\hat{\sigma}_1)
\end{align}
%form the diagram of the KAK decomposition, one commutes the channel of the completed mixed part with the local unitary gate, %and through the commuting, the $\hat{\sigma}_0$ remain the same, but the $\hat{\sigma}_n$(n = 1 or 3) well changing its direction to the linear combination of the Pauli channel, i.e. $\hat{u}({\sigma}_i) = \vec{n}\cdot\vec{\sigma}$.
where $\{\sigma_i\}_i$ is the set of the Pauli operators. \gb{With the consideration} of the KAK decomposition, \gb{if} one commutes the channel of the completely mixed part with the local unitary gates, the $\hat{\sigma}_0$ remains unchanged, but the $\hat{\sigma}_n$ ($ n = 1 $ or $ 3 $) changes direction, becoming a linear combination of the Pauli operator, \textit{i.e.}, $u{\sigma}_iu^\dag = \vec{n}\cdot\vec{\sigma}$.
Under the random sampling, taking the average of all possible single-qubit unitaries, we \gb{obtain},
%Thus after averaging the possible unitary, we had:
\begin{align}
\widetilde{\sigma}_{avg} &= \int_U du\;\hat{u}\circ\hat{\sigma}_i\circ\hat{u}^\dag = \int_{|n|^2=1} dn\;\vec{n}\cdot\vec{\sigma}\otimes\vec{n}\cdot\vec{\sigma}^*
\nonumber\\
&= \frac{1}{3}\sum_{i=1}^3\hat{\sigma}_{i}.
\end{align}
The average EJPP telegating with complete mixed state is then
\begin{align}
  \widetilde{CX}_{avg} &= \iint_{U_a, U_b}du_adu_b\;(\hat{u}_a\otimes\hat{u}_b)\circ\widetilde{CX}\circ(\hat{u}_a^\dag\otimes\hat{u}_b^\dag)\notag\\
  &=\left(\frac{1}{2}\hat{\sigma}_0 + \frac{1}{2}(\frac{\sum_{i=1}^3\hat{\sigma}_i}{4})\right)^{\otimes 2}.
\end{align}
It is equivalent to two single-qubit depolarizing channels with a preserving factor of $1/3$,
\begin{align}
\widetilde{CX}_{avg} &= \widetilde{D}^{(1)}(\frac{1}{3})\otimes \widetilde{D}^{(1)}(\frac{1}{3}),
\end{align}
where $\widetilde{D}^{(1)}(P) = P\hat{\sigma}_0 +\frac{(1-P)}{4}\sum_{i=0}^{3}\hat{\sigma}_i$. \gb{Therefore,} the total error channels should be equal to:
\begin{align}
 \widetilde{\Lambda}(P_e) &= P_e\hat{\sigma}_0^{\otimes 2} + (1-P_e)\;\widetilde{D}^{(1)}(\frac{1}{3})\otimes \widetilde{D}^{(1)}(\frac{1}{3}).
\end{align}
For a large $P_e$, one can approximate the channel by two separable depolarizing channels:
\begin{align}
\widetilde{\Lambda}(P_e)
&\approx \left((\sqrt{P_e}+\frac{1-\sqrt{P_e}}{3})\hat{\sigma}_0+\frac{2(1-\sqrt{P_e})}{3}\widetilde{D}(0)\right)^{\otimes 2}
\nonumber\\
& =\widetilde{D}(\frac{1+2\sqrt{P_e}}{3})\otimes\widetilde{D}(\frac{1+2\sqrt{P_e}}{3})
\end{align}
%And then, the value of $F = P_1P_2 = (1-\epsilon_1)(1-\epsilon_2)$ by the definition, and since we are dealing with the nearly perfect cases( $\epsilon_{1,2}\ll 1\%$) hence:
Here, we assume that a quantum gate acting on auxiliary memory qubits has a sufficiently small error, ${1-P_{a,b}\ll 1\%}$, which leads to a sufficiently small infidelity of the entanglement $1-P_{e} = 1-P_{a}P_{b}\ll 1\%$.
Under this condition, the effective preserving factor of depolarizing channels propagated from noisy gates acting on the auxiliary memory qubits that are involved in the EJPP telegating process can be then approximately given by
\begin{align}
  \frac{1+2\sqrt{P_{e}}}{3} & = \frac{1+2\sqrt{1-(1-P_{a}P_{b})}}{3}
%  \nonumber\\
%  &
  \approx 1-\frac{1}{3}(1-P_{a}P_{b})
  \nonumber \\
  & \approx (P_a P_b)^{\frac{1}{3}}.
\end{align}
It means that the noise of an EJPP telegating channel on two qubits $(q_{1},q_{2})$ introduced by noisy local gates on the auxiliary memory qubits can be effectively described by two single-qubit depolarizing channels acting on the corresponding working qubits with the preserving factor $(P_a P_b)^{\frac{1}{3}}$,
\begin{equation}
  \widetilde{\Lambda}(P_e)
  =
  \widetilde{D}_{q_{1}}(P_a^{\frac{1}{3}} P_b^{\frac{1}{3}})
  \otimes
  \widetilde{D}_{q_{2}}(P_a^{\frac{1}{3}} P_b^{\frac{1}{3}}).
\end{equation}
Note that in this calculation, we do not take the noise in entanglement distribution into account, which is discussed in detail in Appendix \ref{sec::ent_error_ext}.

For an SU(4) gate in the KAK decomposition, there are three CNOT gates. The real implementation of a random SU(4) gate needs three two-qubit gates, each introducing two single-qubit depolarizing channels on each qubit. In total, there should be three single-qubit depolarizing channels on each qubit. In our earlier discussion, we treated the three depolarizing channels induced by the three CNOT gates as one depolarizing channel for our error modeling. Since each CNOT gate in a nonlocal SU(4) gate requires a telegating circuit, originally, there should also be three single-qubit depolarizing channels propagating to the corresponding working qubit originating from the three telegating processes in a SU(4) gate. For consistency, we also treat the three depolarizing channels originating from telegating processes as one depolarizing channel with a preserving factor of $P_{a}^{1/3}P_{b}^{1/3}$.

%Moreover, if one wants to include the effect of non-perfecting of the pre-shared entanglement pairs, it can be done by setting the input entangled state as Eq.(\ref{w_state}) with a parameter $F$, and similarly shifting this error to the working qubit by adding depolarizing channels with error parameter $F^{\frac{1}{3}}$ on each side. For the allocation matrix, we need an additional column for the error cost by the non-perfect entanglement pairs.

\section{Extended cost matrix for EJPP telegating under imperfect entanglement distribution}
\label{sec::ent_error_ext}

\begin{figure*}[htb]
  \centering
  \includegraphics[width=0.9\textwidth]{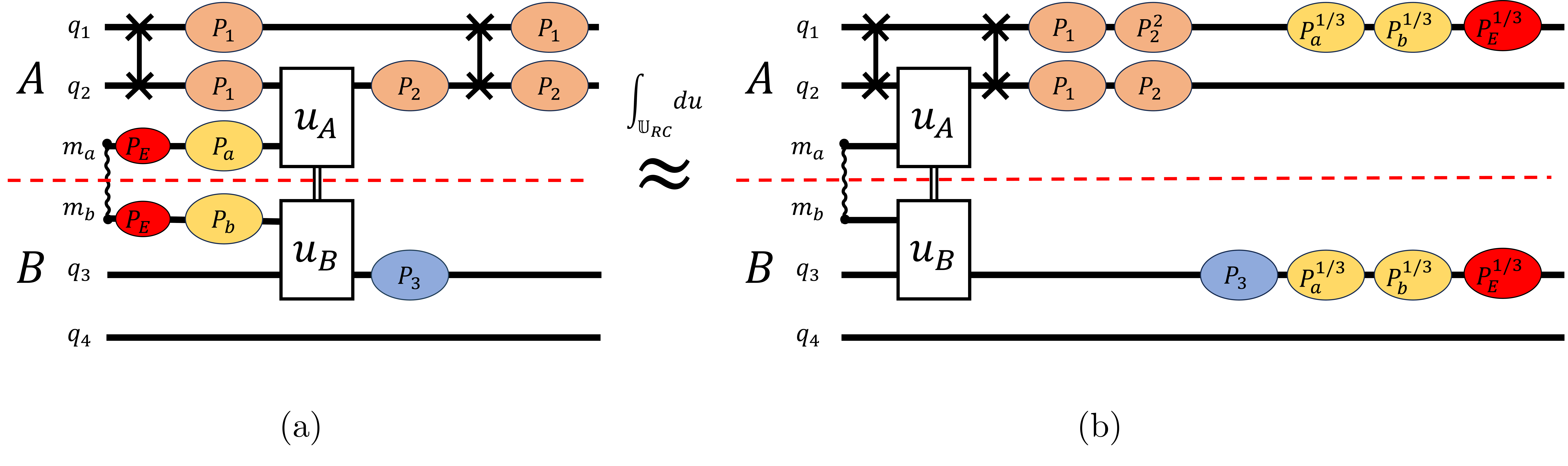}
  \caption{Cost matrix with entanglement noise. The red circles are the depolarizing noises of the entangled pair.
  }\label{fig::cost_mat_ent}
\end{figure*}

To account for the imperfect entanglement distribution, a pre-shared entanglement pairs is expressed as a Werner state similar to Eq.~\eqref{w_state} with an additional factor $P_{E}$
\begin{align}
  \rho_{\Phi} = P_a P_b P_{E}|\Phi^+\rangle\langle\Phi^+|+\frac{1-P_a P_b P_{E}}{4}\mathbb{I}_4.
\end{align}
Employing the same method derived in Appendix \ref{app_exp cost}, one can show that the entanglement distribution error will be transferred to the working qubits by introducing a depolarizing channel on each side with a preserving factor $P_{E}^{\frac{1}{3}}$.
For example, in Fig.~\ref{fig::cost_mat_ent}, the effect of imperfect entanglement is equivalent to adding two depolarizing channels $D(P_{E}^{\frac{1}{3}})$ to the working qubits $q_{2}$ and $q_{3}$, which are then transferred by a swapping gate to the end of $q_{1}$ and $q_{3}$.
Regarding the noise propagation matrix, an additional column representing the error originating from the imperfect entanglement pairs is required to add to the cost matrix.
The cost matrix of the example in Fig.~\ref{fig::cost_mat_ent} is then extended to
\begin{equation}
\label{eq::cost_matrix_eg_ent_noise}
  C(q_{1},q_{3})
  =
  \begin{bmatrix}
    1 & 2 & 0 & 0& \frac{1}{3} & \frac{1}{3} & \frac{1}{3}\\
    1 & 1 & 0 & 0& 0 & 0 & 0\\
    0 & 0 & 1 & 0&  \frac{1}{3} & \frac{1}{3} & \frac{1}{3}\\
    0 & 0 & 0 & 0&  0 & 0 & 0
  \end{bmatrix}.
\end{equation}

\section{Approximation of average gate fidelity in Eq. \eqref{final-eq}}
\label{sec::apdx_approx_AGF}

\begin{figure*}[htb]
    \centering
    \hfill
    \subfloat[Single-QPU devices with size = 6]{\includegraphics[width=0.45\textwidth]{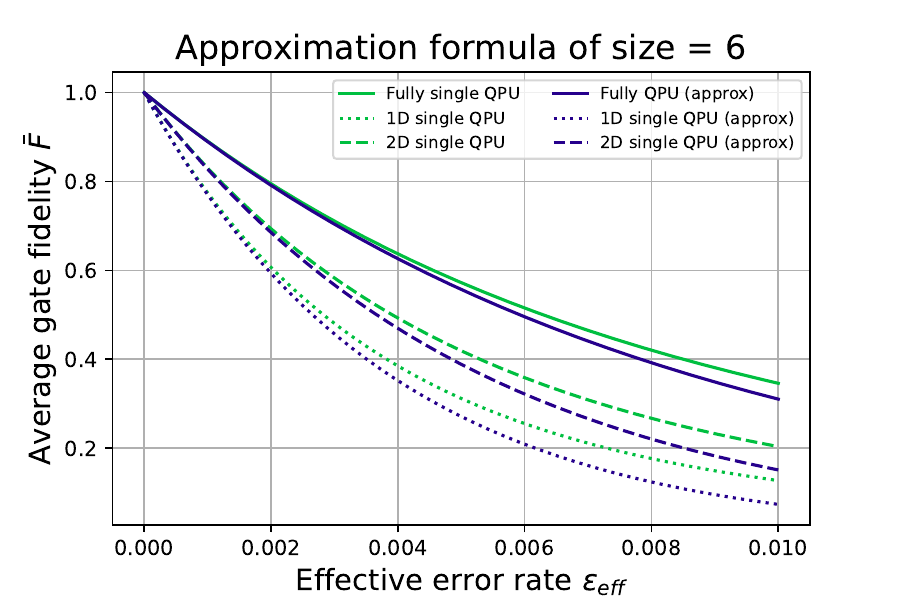}}
    \hfill
    \subfloat[Two-QPU DQC devices with size = 6]{\includegraphics[width=0.45\textwidth]{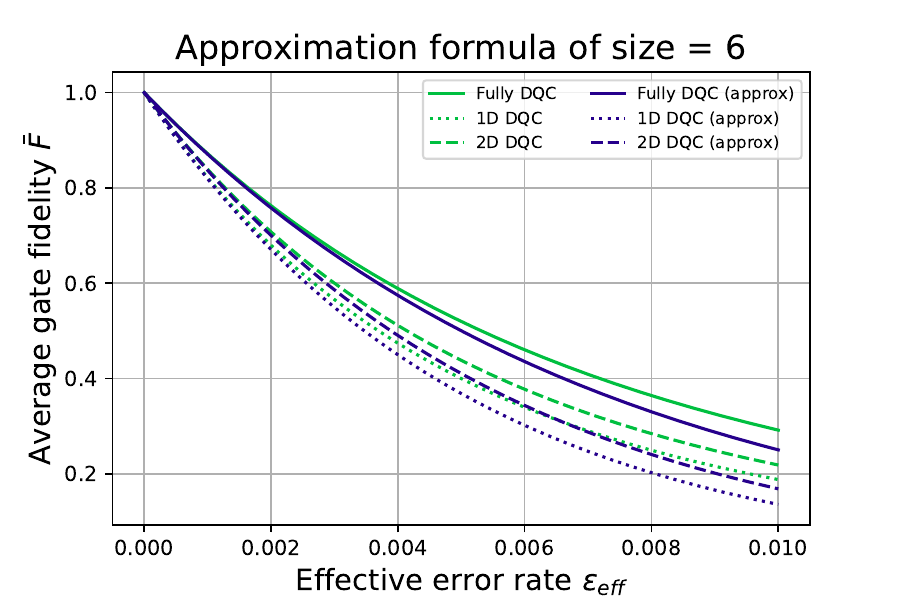}}
    \hfill{ }
    \\
    \hfill
    \subfloat[Single-QPU devices with size = 8]{\includegraphics[width=0.45\textwidth]{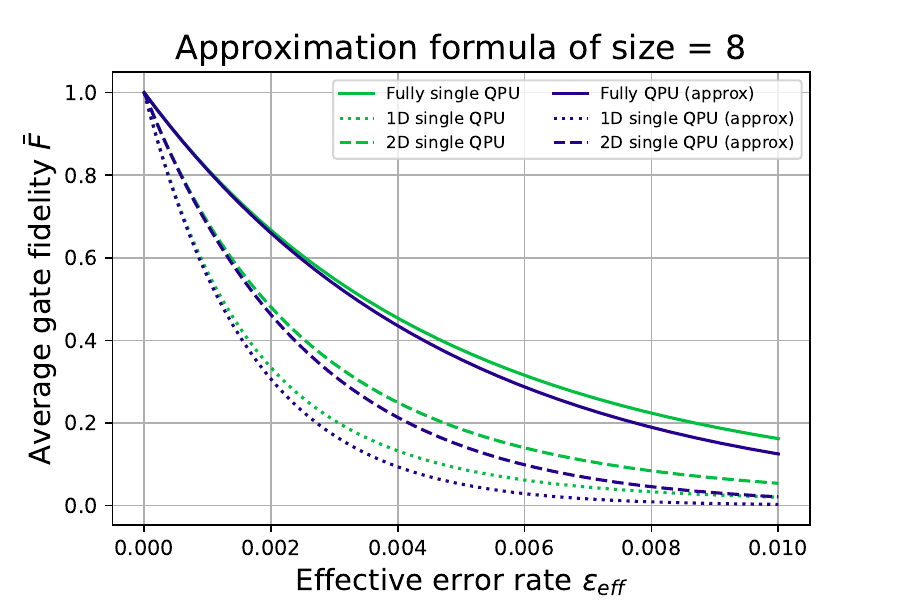}}
    \hfill
    \subfloat[Two-QPU DQC devices with size = 8]{\includegraphics[width=0.45\textwidth]{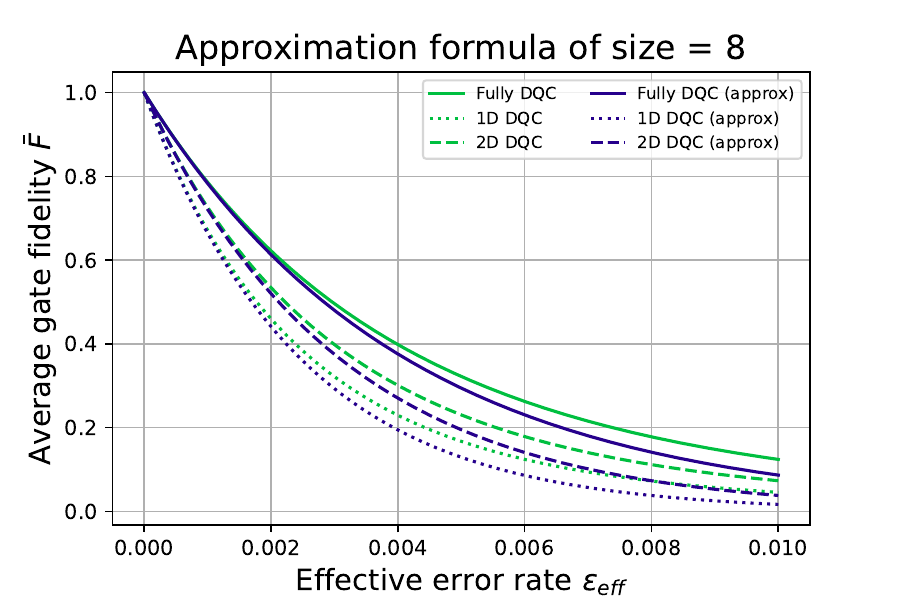}}
    \hfill{ }
    \\
    \hfill
    \subfloat[Single-QPU devices  with size = 10]{\includegraphics[width=0.45\textwidth]{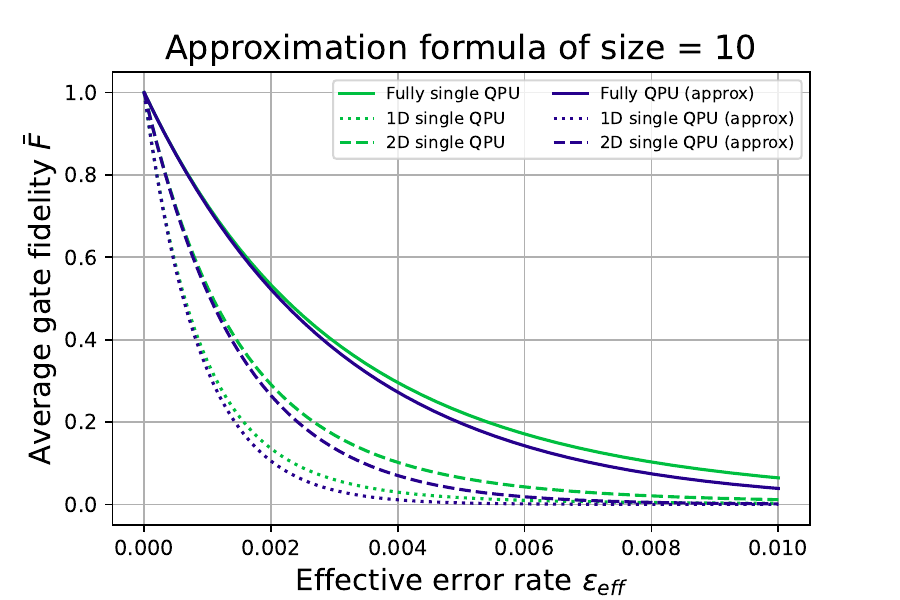}}
    \hfill
    \subfloat[Two-QPU DQC devices with size = 10]{\includegraphics[width=0.45\textwidth]{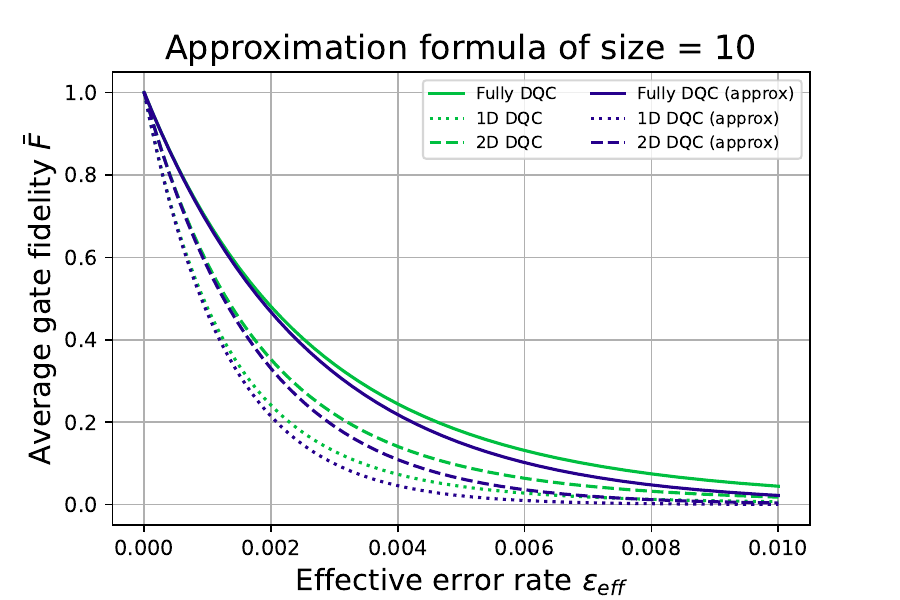}}
    \hfill{ }
    \caption{\wjy{The green curves stand for the original function, and the blue curves stand for the function under approximation. For size 6, although the size is not large, still we had a good match and the high-fidelity region ($\Bar{F}\gtrapprox 0.6$). For the larger size, the approximation gets closer to the original formula.}}
    \label{fig_final-eq}
\end{figure*}

According to Eq. \eqref{modifiy error}, for a small error $\epsilon_{q'}$, the effective preserving factor is approximately given by
\begin{align}
  \bar{P}_{q}(A)
  &= \prod_{q'\in \mathbb{Q}} (1-\epsilon_{q'})^{A_{q,q'}}
  \approx 1-\sum_{q'\in \mathbb{Q}}A_{q,q'}\epsilon_{q'}.
\end{align}
For a large size circuit, we heuristically approximate the preserving factor by its average over the qubits
\begin{equation}
  \bar{P}(A) =
  1- \frac{1}{N}\sum_{q\in\mathbb{Q}_w,q'\in\mathbb{Q}}A_{q,q'}\epsilon_{q'}.
%  \bar{\epsilon} = \frac{1}{N}\sum_{q\in\mathbb{Q}_w}(1-\Bar{P}_q)
%  =\frac{1}{N}\sum_{q\in\mathbb{Q}_w,q'\in\mathbb{Q}}A_{q,q'}\epsilon_{q'}.
\end{equation}
For a large $N$ and a small $\epsilon_{q'}$, the effective preserving factor is approximately given by
\begin{align}
  \wp_{q} & =
  (1- \frac{1}{N}\sum_{q\in\mathbb{Q}_w,q'\in\mathbb{Q}}A_{q,q'}\epsilon_{q'})^{2\lfloor \frac{N}{2} \rfloor}
  \nonumber\\
  & \approx
  1-\sum_{q\in\mathbb{Q}_w,q'\in\mathbb{Q}}A_{q,q'}\epsilon_{q'}.
\end{align}
According to Eq. \eqref{eq::agf_eff_p}, the average gate fidelity is then
\begin{equation}
  \bar{F} \approx \left(1-\frac{1}{2}\sum_{q\in\mathbb{Q}_w,q'\in\mathbb{Q}}A_{q,q'}\epsilon_{q'}\right)^{N}.
\end{equation}
By further assuming that the error rate for each qubit as the same value $\epsilon$, we get a simple formula
\begin{align}
  \Bar{F}_\mathbb{Q}(\epsilon)
  &\approx
  \left(1-\frac{1}{2}\mathcal{A}_{\mathbb{Q}}\epsilon \right)^{N}
  \approx
  \exp(-\frac{1}{2}N\mathcal{A}_\mathbb{Q}\epsilon),
\end{align}
with a characteristic value $\mathcal{A}_\mathbb{Q}\equiv \sum_{q,q'\in \mathbb{Q}_w,\mathbb{Q}}A_{q,q'}$.

%\begin{equation}
%  \Bar{F} \approx \prod_{q\in \mathbb{Q}_w}(1-\frac{N}{2}\braket{\epsilon})\approx \exp(-\frac{N^2}{2}\braket{\epsilon}).
%\end{equation}
%This leads to
%\begin{equation}
%  \Bar{F} = \exp(-\frac{N}{2}\sum_{q,q'\in \mathbf{Q_w},\mathbb{Q}}A_{q,q'}\epsilon_{q'}).
%\end{equation}

  %%%%%%%%%%%%%%%%%%%%%%%%%%%%%%%%%%%%%%%%%%%%%%%%%%%%%%%%%%%%%%%%%%
  %%    Conclusion
%%\section{Conclusion}
%%
  %%%%%%%%%%%%%%%%%%%%%%%%%%%%%%%%%%%%%%%%%%%%%%%%%%%%%%%%%%%%%%%%%%
  %%    Acknowledgements
%%\acknowledgments
%%
  %%%%%%%%%%%%%%%%%%%%%%%%%%%%%%%%%%%%%%%%%%%%%%%%%%%%%%%%%%%%%%%%%%
  %%    Appendix
%%\section{Appendix}
%%                                                                  %%
%%                         Main text                                %%
%%%%%%%%%%%%%%%%%%%%%%%%%%%%%%%%%%%%%%%%%%%%%%%%%%%%%%%%%%%%%%%%%%%%%%

%\newpage
%%%%%%%%%%%%%%%%%%%%%%%%%%%%%%%%%%%%%%%%%%%%%%%%%%%%%%%%%%%%%%%%%%%%%%
%%                         Glossary                                 %%
%%                                                                  %%
\myprintglossary
%%                                                                  %%
%%                         Glossary                                 %%
%%%%%%%%%%%%%%%%%%%%%%%%%%%%%%%%%%%%%%%%%%%%%%%%%%%%%%%%%%%%%%%%%%%%%%

%%%%%%%%%%%%%%%%%%%%%%%%%%%%%%%%%%%%%%%%%%%%%%%%%%%%%%%%%%%%%%%%%%%%%%
%%                         Bibliography                             %%
%%                                                                  %%
\myprintbibliography
%%                                                                  %%
%%                         Bibliography                             %%
%%%%%%%%%%%%%%%%%%%%%%%%%%%%%%%%%%%%%%%%%%%%%%%%%%%%%%%%%%%%%%%%%%%%%%

\end{document}